\documentclass[aps, prl, superscriptaddress, twocolumn, showpacs]{revtex4-1}
\usepackage[utf8]{inputenc}
\usepackage{bm}
\usepackage{mathtools}
\usepackage{bbold}
\usepackage{amsmath}
\usepackage{amssymb}
\usepackage{color}
\usepackage[dvipsnames]{xcolor}
\usepackage{multirow}
\usepackage{comment}
\usepackage{cancel}
\usepackage{float}
\usepackage{color}
\usepackage[dvipsnames]{xcolor}
\usepackage{epstopdf}
\usepackage{amsmath}
\usepackage{braket}
\usepackage{amsthm}
\usepackage{amssymb}
\usepackage{amsfonts}
\usepackage{graphicx,subfigure}
\usepackage{txfonts}
\usepackage{ulem}
\usepackage{mathtools}
\usepackage{bm}
\usepackage{bbm}
\usepackage{hyperref}
\usepackage{natbib}
\newcommand{\vect}[1]{\boldsymbol{#1}}

\begin{document}

\newcommand{\INO}{Istituto Nazionale di Ottica, Consiglio Nazionale delle Ricerche (INO-CNR), Largo Enrico Fermi 6, 50125 Firenze, Italy}
\newcommand{\LENS}{European Laboratory for Nonlinear Spectroscopy (LENS), Via N. Carrara 1, 50019 Sesto Fiorentino, Italy}
\newcommand{\UNIFI}{Physics Department, University of Florence, Via G. Sansone 1, 50019 Sesto Fiorentino FI, Italia
}
\newcommand{\UNINA}{Physics Department, University of Naples Federico II, Via Cintia 21, 80126 Naples, Italy}
\newcommand{\QSTAR}{QSTAR, Largo Enrico Fermi 2, 50125 Firenze, Italy}
\newcommand{\be}{\begin{equation}}
\newcommand{\ee}{\end{equation}}

%\title{Analogue many-body gravitating quantum systems with a network of \\dipolar Bose–Einstein condensates}

\title{Many-body gravitating quantum systems \\ with Bose–Einstein condensates and dipolar analogue}

\author{Youssef Trifa}
%\email{youssef.trifa@ino.cnr.it}
\affiliation{\INO}
%\affiliation{\LENS}
%\affiliation{\QSTAR}
\author{Dario Cafasso}
\affiliation{\INO}
\affiliation{\UNINA}
\author{Marco Fattori}
\affiliation{\INO}
\affiliation{\LENS}
\affiliation{\UNIFI}
%\author{Augusto Smerzi}
%\affiliation{\INO}
%\affiliation{\LENS}
%\affiliation{\QSTAR}
\author{Luca Pezz\`e}
%\email{luca.pezze@ino.cnr.it}
\affiliation{\INO}
\affiliation{\LENS}
%\affiliation{\QSTAR}
\date{\today}
	
\begin{abstract}

Quantum probes of gravity in the Newtonian regime, based on mass–energy equivalence in clocks or spatial superpositions in interferometers, share a 
common description in terms of an effective qubit-qubit coupling. 
Here we extend this framework to atomic ensembles, regarded as interacting collective qudits. 
The many-body enhancement boosts the signal-to-noise and increases the effective interaction rate, facilitating the observation of gravitationally-induced entanglement and decoherence, certified by metrological witnesses based on local and collective spin squeezing.
We further identify trapped bimodal Bose–Einstein condensates with long-range interactions, including dipolar couplings, as a programmable analogue platform for simulating gravitating quantum dynamics at accessible time and energy scales. 
Extending the protocol to a sensor network broadens the entanglement-detection window.
\end{abstract}

\maketitle

\section{Introduction}

%
%Testing whether the gravitational field obeys quantum superposition using a Stern-Gerlach experiment is a compelling idea that traces back to Feynman in 1957~\cite{Feynman57}, while conventional quantum gravity scenarios have traditionally been associated with the need to probe extremely high energy scales~\cite{Maggiore2005, Carrol2019, Isham1992, Oriti2009, Dyson2013}. 
%
Traditionally, the search for quantum gravity has been associated with inaccessible high-energy scales~\cite{Maggiore2005, Carrol2019, Isham1992, Oriti2009, Dyson2013}. 
However, Feynman proposed an alternative path as early as 1957: testing whether the gravitational field can exist in a quantum superposition using a Stern-Gerlach setup~\cite{Feynman57}. He argued that coupling a quantum superposition to a strictly classical gravitational field would break quantum coherence -- as theorized in modern collapse frameworks~\cite{Ghirardi1986, Diosi1989, Penrose1996, Howl2019} -- or yield fundamental physical inconsistencies, thereby implying that the superposition principle must also apply to gravity.
Recent advancements in quantum technology have further refined Feynman's idea, paving the way for novel quantum-information-inspired approaches to seek indirect signatures of gravity's non-classical nature within low-energy regimes~\cite{GiacominiQUANTUM2021, Belenchia2018, BoseRMP2025, MarlettoRMP2025, Bose2017, Marletto2017, MarlettoPRD2020, Zych2011, PikovskiNATPHYS2015, Castro-RuizPNAS2017, Castro-Ruiz2020}.
%
%\think{Testing the nonclassical nature of gravity is traditionally associated with the need to probe extremely high energies, near the Planck scale, well beyond present experimental capabilities~\cite{Maggiore2005, Carrol2019, Isham1992, Oriti2009, Dyson2013}. 
%
%In contrast, recent quantum-information-inspired approaches seek indirect signatures of such nonclassical features in low-energy regimes~\cite{BoseRMP2025, MarlettoRMP2025, Bose2017,Marletto2017,MarlettoPRD2020,Castro-RuizPNAS2017,Castro-Ruiz2020, PikovskiNATPHYS2015}.
%
%Two paradigmatic gedanken experiments exemplify this idea.}

Among the frameworks that preserve the structure of quantum mechanics, 
%without invoking a full theory of quantum gravity, 
two paradigmatic gedanken experiments exemplify this idea.
In the Bose--Marletto--Vedral (BMV) proposal~\cite{Bose2017, Marletto2017, MarlettoPRD2020}, two masses are each prepared in a spatial superposition, e.g. placed in the arms of two spatial interferometers.
Depending on its path, each mass generates a different Newtonian interaction for the mass in the other interferometer.
In the clock-based proposal of Castro-Ruiz, Giacomini and Brukner (CGB)~\cite{PikovskiNATPHYS2015,Castro-RuizPNAS2017,Castro-Ruiz2020,Belenchia2018}, the effect of gravitational interaction is due to the mass-energy equivalence principle for two spatially-separated quantum clocks.
%
%\think{This also implies fundamental limitations to the measurability of time~\cite{Chen2025, PikovskiNATPHYS2015, Castro-RuizPNAS2017, Castro-Ruiz2020, PaigePRL2020, SmithNATCOMM2020, GiacominiQUANTUM2021, CafassoPRD2024}.}
%
Consequently, each clock experiences a different quantum time-dilation depending on the internal energy of the others ~\cite{Zych2011, PikovskiNATPHYS2015, Castro-RuizPNAS2017, Belenchia2018, Castro-Ruiz2020, PaigePRL2020, SmithNATCOMM2020, GiacominiQUANTUM2021, CafassoPRD2024}.

Both scenarios address gravity's non-classicality by observing effects related to gravitationally-induced entanglement (GIE) and gravitationally-induced decoherence (GID). In these proposals, decoherence arises from the partial trace of a unitary entangling dynamics, rather than a fundamental modification of quantum mechanics~\cite{Howl2019,PikovskiNATPHYS2015}.
However, probing GIE and GID is challenging~\cite{MarshmanPRA2020, OverstreetPRD2023, KrisnandaNPJQI2020, OppenheimNATCOMM2023, AngeliARXIV}: besides requiring large masses, short distances and long integration times,
many current proposals effectively operate at the single-qubit level, yielding limited information per experimental shot.
In contrast, atomic ensembles~\cite{HaineNJP2021, CarneyPRXquantum2021} offer several key advantages thanks to the well controlled coherent manipulation~\cite{ChinRMP2010, CroninRMP2009, PezzeRMP2018} of internal and external degrees of freedom, the well controlled separation of the clouds~\cite{KovachyNATURE2015, AsenbaumPRL2017}, and the long trapping times~\cite{XuSCIENCE2019}: these properties have made atomic systems central for classical gravity measurements~\cite{RosiNATURE2014, BothwellNATURE2022} and fundamental tests at the quantum interface~\cite{Loriani2019, OverstreetSCIENCE2022, Fromonteil2025, Paczos2025}.
Furthermore, devising strategies to overcome practical limitation motivates analogue-gravity platforms as testbeds to benchmark protocols and develop experimentally viable witnesses.
While analogue systems have successfully addressed aspects of black-hole physics~\cite{UnruhPRL1981, VisserCQG1998, BarceloLRR2011, Mannarelli2021, Poli23, BraunsteinNRP2023}, dedicated simulators of quantum-gravity signatures remain scarce~\cite{Polino2024, Bian2025}.

%%%%%%%%%%%%%%%%%%%%%%%%%%%%%%%%%%%%%%
%% FIGURE 1
%%%%%%%%%%%%%%%%%%%%%%%%%%%%%%%%%%%%%%
\begin{figure}[b!]
\includegraphics[width=0.42\textwidth]{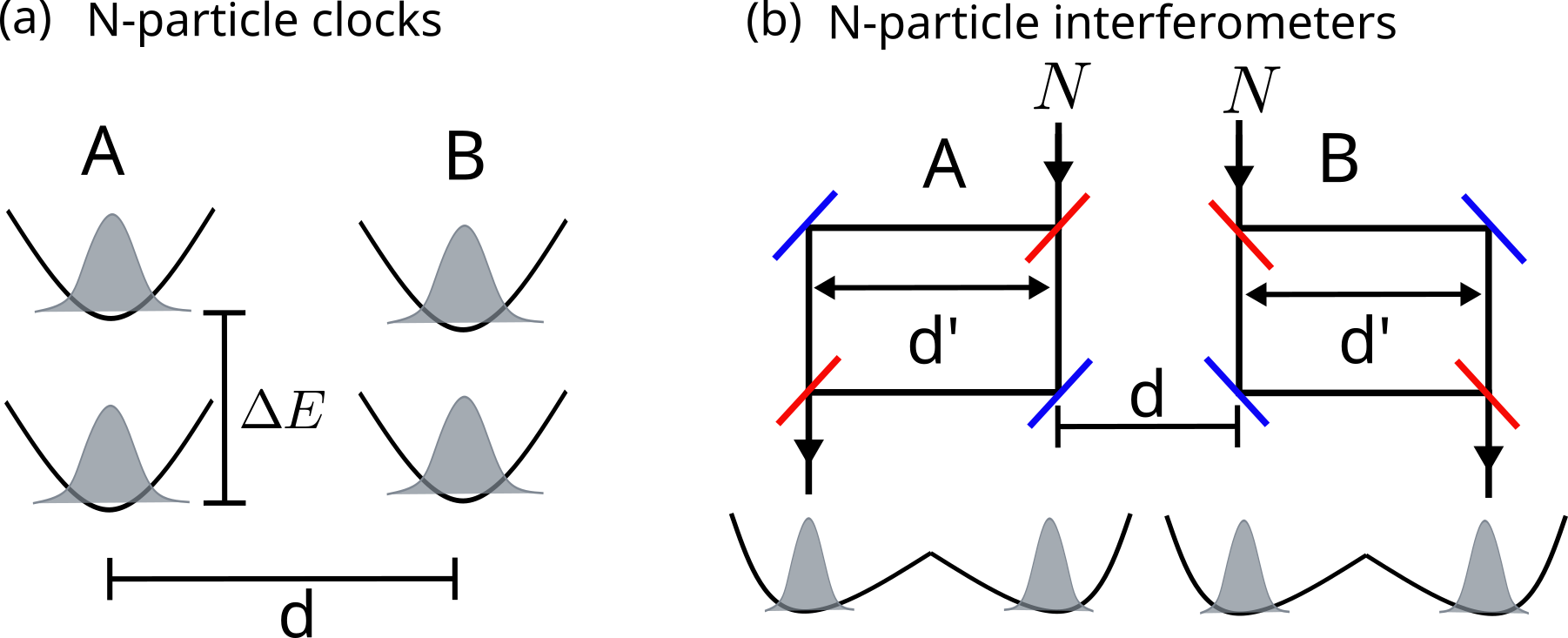}
\hspace{1mm}
\caption{{\bf Generalization of the CGB and BMV proposals to trapped bimodal BECs.}
(a) Schematic of two quantum clocks, labeled as A and B, each made of $N$-particle bimodal BEC placed at a distance $d$. 
The two modes are two internal levels, with energy gap $\Delta E$. 
(b) Schematic of two $N$-particle path atom  interferometers (top).
The two nearby arms are separated by a distance $d$ while the inter-arm separation is $d'$.
Beam-splitters are shown in red and mirrors in blue.
This configuration can be realized with BECs in two double-well potentials (bottom), each well acting as an external mode.
}
\label{Fig0}
\end{figure}
%%%%%%%%%%%%%%%%%%%%%%%%%%%%%%%%%%%%%%
%%%%%%%%%%%%%%%%%%%%%%%%%%%%%%%%%%%%%%
%%%%%%%%%%%%%%%%%%%%%%%%%%%%%%%%%%%%%%

Here we generalize both the CGB and the BMV proposals to the case of bimodal Bose-Einstein condensates (BECs).
Spatially-separated condensates can be put in a superposition of two internal states, Fig.~\ref{Fig0}(a), realizing  composite gravitating quantum clocks.
Alternatively, the two condensates can be put in a spatial superposition, realizing many-body gravitating interferometers, Fig.~\ref{Fig0}(b).
%
%with a large number of particles, $N$, thus realizing  composite gravitating quantum clocks, Fig.~\ref{Fig0}(a), and \dario{many-body gravitating} interferometers, Fig.~\ref{Fig0}(b).
%
We show that both settings are governed by the same gravitational 
%interaction 
Hamiltonian, 
that includes both local (in each condensate) and non-local (among condensates) couplings. 
The former can be canceled out by tuning the particle-particle scattering length, thus reducing the gravitational interaction to a genuine 
%\del{gravity-induced} 
qudit–qudit coupling.
The many-body probe boosts the signal-to-noise ratio of existing single-qubit proposals and enables state-engineered enhancements that accelerate by a factor $\sim N$ the time at which 
%detection of 
GIE and GID 
%dynamics
effects arise.
%
%\think{Moreover, the gravitational interaction can be emulated by using long-range (e.g. dipolar) coupling between the atoms, paving new ways to engineering GIE and GID analogue dynamics in current ultracold-atom experiments.}

Moreover, the same interaction Hamiltonian can be implemented in current ultracold-atom experiments by emulating the gravitational interaction via dipolar coupling. This provides a robust and highly sensitive platform for engineering analogue dynamics of gravitationally induced entanglement (GIE) and gravitationally induced decoherence (GID).
Such an analogue implementation can inform about experimental aspects that are intractable from a computational point of view in the large-$N$ limit, such as the impact of internal (inside each condensate) decoherence and external noise on entanglement characterization protocols. This would also help calibrate future experiments and devise noise mitigation schemes that are compatible with GIE and GID detection.
%—that remain computationally intractable to simulate numerically in the presence of noise?
Furthermore, the accurate level of control reached by these platforms provides a versatile playground to investigate quantum time-dilation effects in composite quantum clocks, offering an experimental testbed for exploring the low-energy interface of quantum mechanics and relativity.
Finally, we show that extending this setup to small networks of BECs further enhances the sensitivity to these gravitational signatures.

\section{Results}

{\bf Gravitating quantum systems.}
In the Newtonian regime, the gravitational interaction between two particles of mass $m_A$ and $m_B$ respectively, placed at a relative distance $d$, is described by Newton's potential $U(d) = -G m_A m_B/d$, where $G$ is the gravitational constant. 
Invoking mass--energy equivalence, the interaction also depends on the internal energy of each particle, e.g. modeled as a qubit by CGB~\cite{Castro-RuizPNAS2017}.
This effect is described by replacing the classical mass with a mass operator for each system, $m_j \to  m_{j} \openone + (\Delta E/c^2) \hat{\sigma}^z_j$, where $c$ is the speed of light, the Pauli matrix $\hat{\sigma}^z_j$ describes energy excitations for the particle $j=A,B$, and $\Delta E$ is the energy gap between the two internal levels.
This yields a quantum version of Newton's potential, $U(d) \to  \chi^G_{\rm CGB}  \, \hat{\sigma}^z_A \otimes \hat{\sigma}^z_B$, up to irrelevant local terms, with coupling constant $\chi^{G}_{\rm CGB} = -G \Delta E^{2} /(d c^4 )$. 

On the other hand, the BMV proposal relies on creating a nonlocal superposition of two massive particles~\cite{Bose2017, Marletto2017} using two interferometers with spatially separated arms. 
Identifying the two paths with a qubit basis, the branch-dependent Newtonian phases lead to an effective interaction that is proportional to $\chi^{G}_{BMV} \hat{\sigma}^z_A \otimes \hat{\sigma}^z_B$,  with $\chi^{G}_{BMV} = -G m^2 /d$, where $m$ is the mass of a single particle, and $d \ll d'$ are the distances between nearby and inner arms of each interferometer.
Thus, the core mechanism behind the CGB and BMV proposals is a qubit-qubit 
%Ising 
coupling 
%described by the 
in the form $\hat{\sigma}^z_A \otimes \hat{\sigma}^z_B$
%term
.

We generalize both proposals by considering two trapped and spatially separated bimodal BECs, each containing $N$ bosons.
The two modes represent either internal states with energy spacing $\Delta E$ (CGB generalization), Fig.~\ref{Fig0}(a), or two external interferometer arms (BMV), Fig.~\ref{Fig0}(b).
In both cases the dynamics is described by the Hamiltonian
\begin{equation} \label{eq:HGbec}
    \hat{H}_{\rm G} = (\chi_{\rm loc}^G+\chi_{\rm cont})  \big(  (\hat{J}^z_A)^{2} + (\hat{J}^z_B)^{2} \big) + \chi^G_{\rm nloc} ~\hat{J}^z_A \otimes \hat{J}^z_B,
\end{equation}
where $\hat{J}^x_{j}$, $\hat{J}^y_{j}$ and $\hat{J}^z_{j}$ are collective pseudo-spin operators for an effective qudit $j=A,B$ of dimension $N+1$ (see Methods).
The first term includes the gravitational ($\chi_{\rm loc}^G$) plus the contact ($\chi_{\rm cont}$) contribution within each BEC, and $\chi_{\rm nloc}^G$ describes the nonlocal gravitational coupling between the spatially-separated condensates.
Local terms can be tuned and, if desired, canceled by using a magnetic field near a Feshbach resonance~\cite{ChinRMP2010}.
In this case, the $N$-particle Hamiltonian~\eqref{eq:HGbec} becomes $\hat{H}_{G} = \chi^G_{\rm nloc}~\hat{J}^{z}_{A}\otimes \hat{J}^{z}_{B}$.
Furthermore, since $(\hat{\sigma}^z_j)^2 = \openone$, the simple qubit-qubit interaction is recovered when $N=1$.
%Notice that, using $(\hat{\sigma}^z_j)^2 = \openone$, the Hamiltonian $\hat{H}_{G}$ reduces to $\sim \hat{\sigma}^z_A \otimes \hat{\sigma}^z_B$ for $N=1$.

We give in Section~A and~B of Supplemental Material a derivation of Eq.~(\ref{eq:HGbec}) in the CGB case 
%(where, in particular $\chi^{G}_{\rm nloc} = \chi^{G}_{\rm CGB}$) 
and the BMV case, respectively.
As a consequence of the mass-energy equivalence principle, this equation
%Eq.~(\ref{eq:HGbec}) 
is also consistent with the gravitational Hamiltonian derived in Ref.~\cite{Chen2025} for delocalized quantum sources. 
In particular, in the absence of a whole theory of quantum gravity, the latter derivation ensures that the BEC mapping is well-defined at low energies -- within the framework of linearized quantum gravity.

Furthermore, the gravitational interaction in Eq.~(\ref{eq:HGbec}) 
can be emulated by trapped bimodal BECs with long-range, e.g. dipolar, interactions~\cite{LahayeRPP2009, DefenuRMP2023, ChomazRPP2023, bigagliNat2024}. 
This %\think{analogue} 
platform provides 
%exactly 
the same spin-spin Hamiltonian 
%as Eq.~\eqref{eq:HGbec}, 
with $\chi^G_{\rm loc}$ and $\chi^G_{\rm nloc}$ replaced by the coupling terms $\chi_{\rm loc}$ and $\chi_{\rm nloc}$ due to the long-range interaction within each BEC and between separated condensates, respectively, see Section C of Supplemental Material for a microscopic derivation of these coefficients.
This leads to $\chi_{\rm nloc}\sim 1/d^{\alpha}$ with, for instance,
$\alpha=3$ for dipolar interaction, whereas gravity gives $\chi_{\rm nloc}^G\propto 1/d$.
For fixed spatial separation between the BECs, this extra spatial dependence in the analogue system can be absorbed into an effective parameter $g_{\rm eff}$ such that $\chi_{\rm nloc} \sim g_{\rm eff} /d$, hence enabling a faithful analogue of the gravitational interaction at a relative distance $d$.

Overall, Eq.~\eqref{eq:HGbec} accounts for the superposition of both spatial locations and internal energy states, providing a common low-energy description of the physics at the intersection of our most fundamental theories. This description naturally follows from the the quantum superposition principle and mass-energy equivalence, thereby enabling investigation of their compatibility by means of current quantum technologies.

%%%%%%%%%%%%%%%%%%%%%%%%%%%%%%%%%%%%%%
%% FIGURE 2
%%%%%%%%%%%%%%%%%%%%%%%%%%%%%%%%%%%%%%
\begin{figure}[t!]
\includegraphics[width=0.41\textwidth]{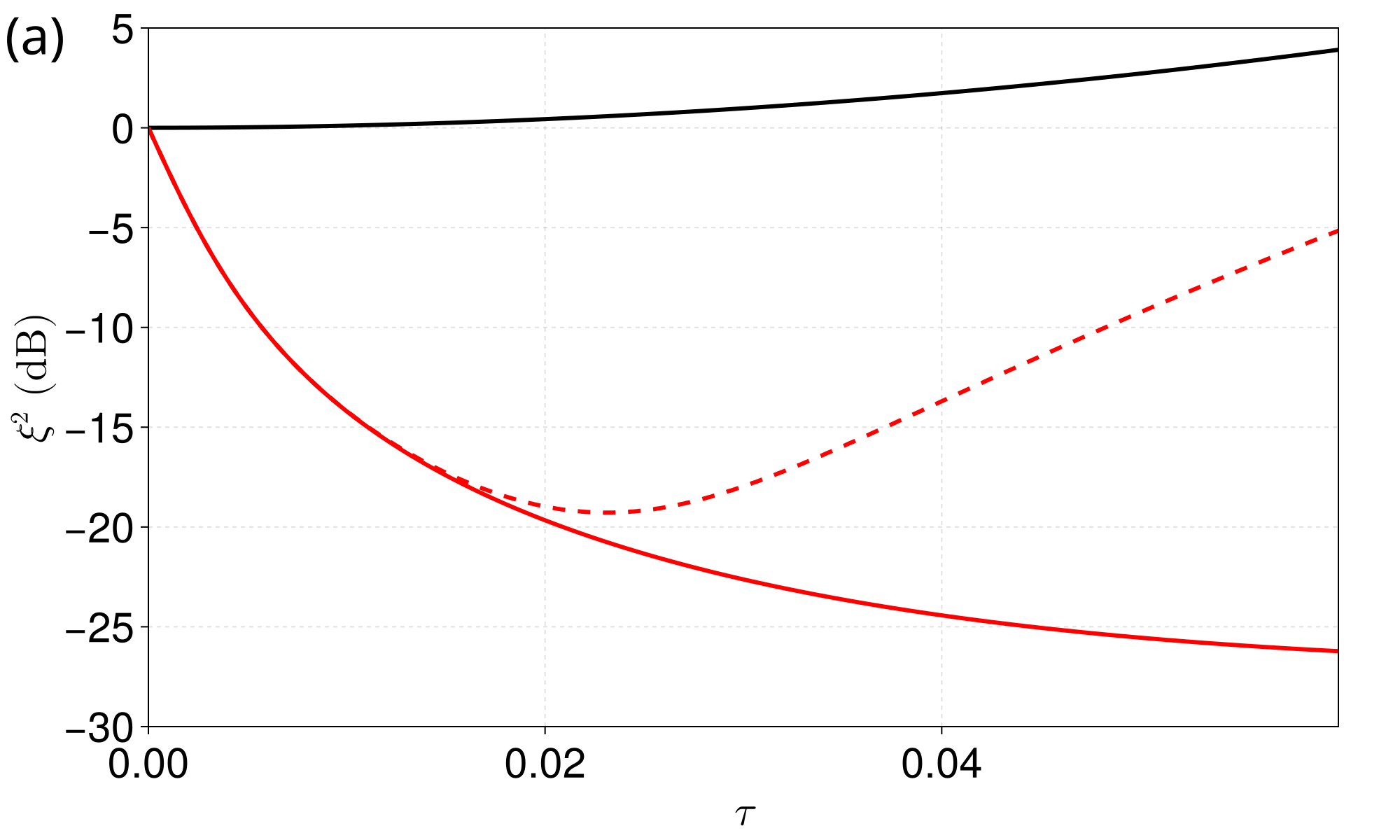}
\includegraphics[width=0.41\textwidth]{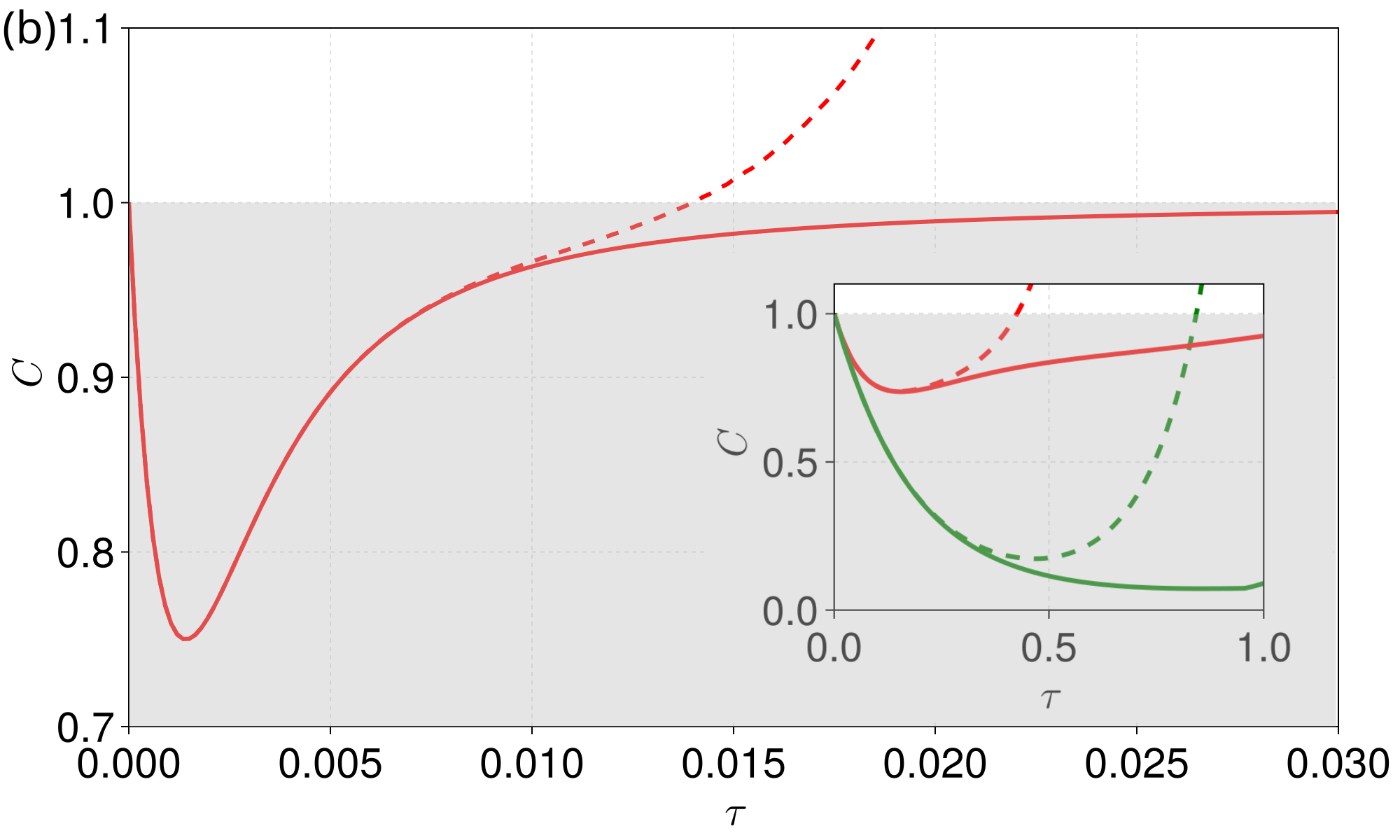}
\caption{
{\bf GIE dynamics.}
We start with all the atoms in the same mode for each BEC.
At $t=0$ a $\pi/2$ pulse distributes the particles among the two modes. 
The subsequent dynamics is a free evolution according to the Hamiltonian $J_A^z\otimes J_B^z$.
(a) Local ($\xi^{2}_{\rm loc}$, solid black) and collective ($\xi^{2}_{\rm col}$, dashed red) spin squeezing and inverse Fisher information ($2N/F_{\rm col}$ solid red line) as a function of the effective time $\tau =\chi_{\rm nloc}^G t$. % 
From these quantities we compute the coefficients ${\cal C}_{1}$ and ${\cal C}_{2}$, shown as red solid line and red dashed line in panel (b), respectively. 
Accessing the gray region (${\cal C}_{1}, {\cal C}_{2} < 1$) witnesses entanglement between the two ensembles.
The inset shows ${\cal C}_{1}$ (red solid) and ${\cal C}_{2}$ (red dashed) as well as a tighter version of these criteria, $\tilde{{\cal C}}_{1}$ (green solid) and $\tilde{{\cal C}}_{2}$ (green dashed), based on the local Fisher information matrix instead of the local covariance matrix, see Methods for more details.
For both figures, we have taken $N = 1000$ for the main panels, and $N=10$ for the inset of figure (b).
Results shown in the main panels of figures (a-b) are fully analytical and obtained following Ref.~\cite{KurkjianPRA2013}, see details in Supplemental Material Section F.
}
\label{Fig1}
\end{figure}
%%%%%%%%%%%%%%%%%%%%%%%%%%%%%%%%%%%%%%
%%%%%%%%%%%%%%%%%%%%%%%%%%%%%%%%%%%%%%
%%%%%%%%%%%%%%%%%%%%%%%%%%%%%%%%%%%%%%

{\bf Many-body GIE.}
%
%\youssef{Forse è meglio di parlare qui di "many-body GIE"  per il titolo, per questa parte è vera per l'analogue ma anche per la gravità con many qubits (solo cambia il prefactor per il tempo). Stessa cosa per la parte Analogue GID $\rightarrow$ Many-body GID ? }
%
Our first protocol targets {many-body} GIE dynamics.
With the possibility to tune the contact interaction, we set the net local nonlinearity 
$\chi_{\rm loc}^{G}+\chi_{\rm cont}$ to $0$, and we introduce the rescaled evolution time $\tau\equiv \chi_{\rm nloc}^G t$.
Each BEC is initially prepared with all atoms in a single mode.
The corresponding initial product state is an eigenstate of $\hat J_A^z\otimes \hat J_B^z$ and therefore evolves only by an overall phase.
The dynamics starts at $\tau=0$, when we apply a $\pi/2$ rotation,
e.g. $e^{-i (\pi/2) \hat{J}^y_A} \otimes e^{-i (\pi/2) \hat{J}^y_B}$, thereby preparing local coherent spin states aligned along the $x$ axis, $|\mathrm{CSS}_x\rangle_A\otimes|\mathrm{CSS}_x\rangle_B$.
The rotation can be implemented either by a Rabi pulse coupling internal energy levels, or by controlling the tunneling dynamics in double-well case~\cite{PetruccianiNATCOMM2026}. 
We then let evolve the system for time $\tau$ according to $\ket{\psi(\tau)} = e^{-i \tau \hat{J}^{z}_{A}\otimes\hat{J}^{z}_{B}} \ket{CSS_x}_A\otimes \ket{CSS_x}_B$.

To witness entanglement between the two ensembles, we monitor the following two quantities:
\begin{equation}\label{ent_crit}
    {\cal C}_{1} \equiv \frac{8 \Gamma_{\rm loc}}{F_{\rm col}}, \quad {\rm and} \quad 
    {\cal C}_{2} \equiv \frac{ 4 \xi^2_{\rm col} \Gamma_{\rm loc}}{N}.
    \,
\end{equation}
Here, $\Gamma_{\rm loc}$ is the largest eigenvalue of the covariance matrix of local pseudo-spin operators, $F_{\rm col}$ is the quantum Fisher information optimized over collective spin directions, and $\xi^2_{\rm col}$ is the Wineland spin-squeezing parameter minimized over orthogonal collective quadratures: see Methods for definitions and details.
The quantity ${\cal C}_{2}$ can be experimentally accessed by measuring first and second moments of collective spin operators.
For separable states, one has ${\cal C}_{1}\geq 1$~\cite{GessnerPRA2016, LiPRA2013} and ${\cal C}_{2}\geq 1$,
the latter following from $\xi_{\rm col}^2\ge 2N/F_{\rm col}$.
Thus, ${\cal C}_{1}<1$ or ${\cal C}_{2}<1$ certify entanglement between ensembles $A$ and $B$ without any assumption about the purity on the purity of states.
Typically ${\cal C}_{1}\le {\cal C}_{2}$, while the two quantities coincide for spin-Gaussian states. 
Other bipartite witnesses are also possible, including criteria based on Einstein--Podolsky--Rosen correlations in BECs~\cite{KurkjianPRA2013, ByrnesPRA2013, PeiseNATCOMM2015, FadelSCIENCE2018, KunkelSCIENCE2018, LangeSCIENCE2018}.

As shown in Fig.~\ref{Fig1}(a), the nonlocal coupling generates collective spin squeezing ($\xi_{\rm col}^2<1$), without producing local squeezing ($\xi_{\rm loc}^2\ge 1$), consistent with correlations building exclusively between the two subsystems.
Equivalently, the collective squeezing originates exclusively from inter-ensemble correlations, i.e.\ from off-diagonal blocks of the covariance (squeezing) matrix (see Methods), thus providing a powerful witness of GIE in the many-body system. 
At early times the collective squeezing saturates the Fisher bound,
$\xi_{\rm col}^2=2N/F_{\rm col}$.
This is also visible in Fig.~\ref{Fig1}(b), where the two criteria coincide at short times (e.g.\ up to $\tau\lesssim 5\times 10^{-3}$ for $N=1000$). 
The witness ${\cal C}_{2}$ therefore detects bipartite entanglement at short times, whereas ${\cal C}_{1}$ remains effective at all times, at the price of requiring increasing precision to resolve deviations from unity as $\tau$ grows.
The minimum value ${\cal C}_{1}\approx {\cal C}_{2}\approx 0.75$ is essentially independent of $N$ and is reached at $\tau\sim 1/N$.
Crucially, increasing $N$ shifts the entanglement-detection window to shorter times, enabling earlier observation of GIE, see inset of Fig.~\ref{Fig1}(b) for a plot of the case $N=10$.

%%%%%%%%%%%%%%%%%%%%%%%%%%%%%%%%%%%%%%
%% FIGURE 3
%%%%%%%%%%%%%%%%%%%%%%%%%%%%%%%%%%%%%%
\begin{figure}[ht!]
\includegraphics[width=0.45\textwidth]{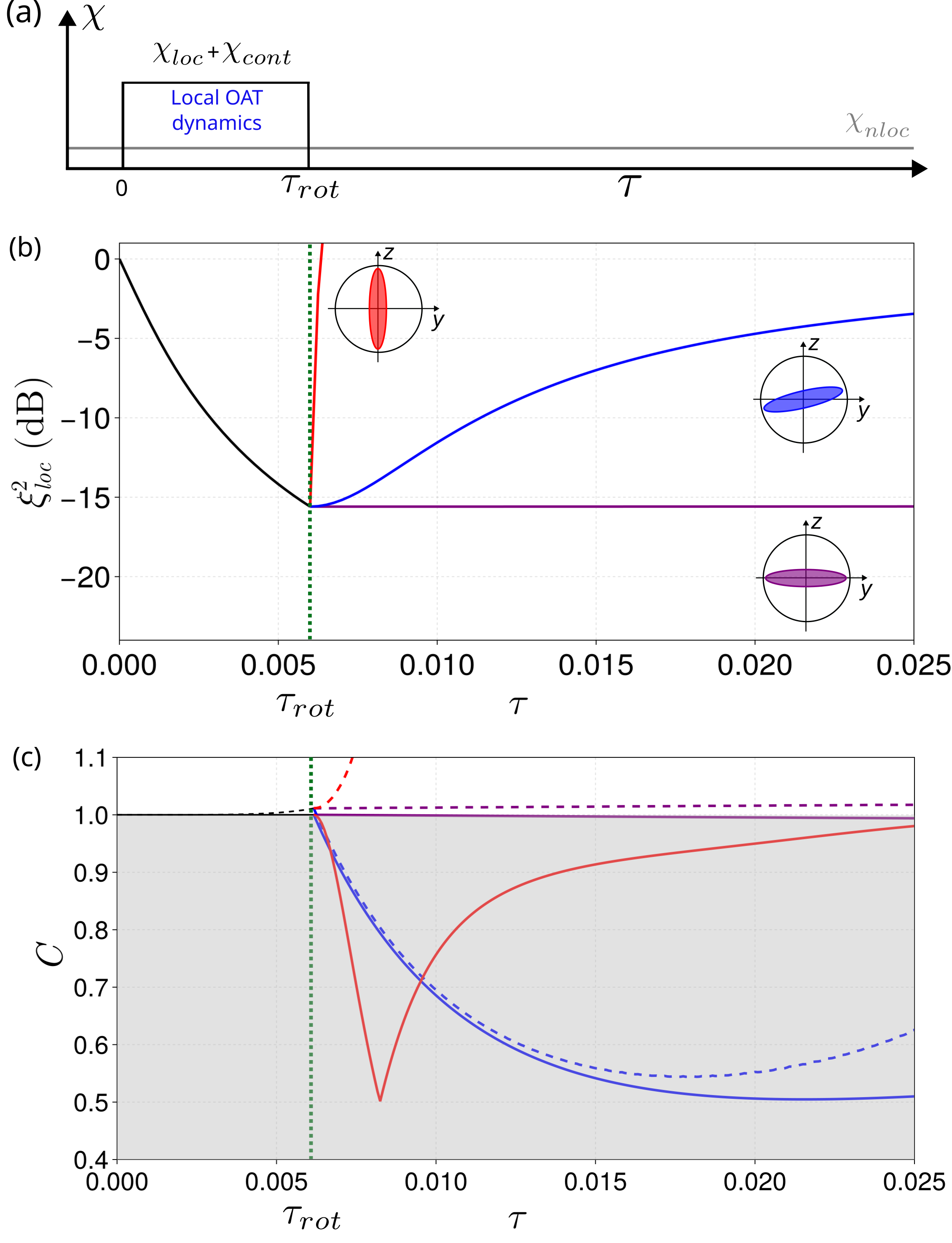}
\caption{{\bf GID dynamics.}
(a) Schematic of the experimental sequence. 
The local interaction is turned on at time $t=0$, to generate spin squeezing in each ensemble.
It is turned off at $t=t_{\rm rot}$, when an additional local rotation of the state by an angle $\theta$ is considered.
The subsequent dynamics is a free evolution coupling the two ensembles.
(b) Local spin squeezing as a function of the effective time $\tau$. 
In particular, $\tau = (\chi_{\rm loc}^G + \chi_{\rm cont}) t$ for $0 \leq t \leq t_{\rm rot}$ and $\tau = (\chi_{\rm loc}^G + \chi_{\rm cont}) t_{\rm rot} + \chi_{\rm nloc}^G (t-t_{\rm rot})$ for $t \geq t_{\rm rot}$. 
The black solid line corresponds to the spin squeezing obtained before $\tau_{\rm rot}$. 
The different colored lines, starting from $\tau_{\rm rot}$, correspond to the different orientations of the local spin squeezing: $\beta = 0$ (purple), $\beta = \pi/24$ (blue), and $\beta = \pi/2$ (red). 
The inset shows the corresponding state on the Bloch sphere after the rotation at time $\tau_{\rm rot}$.
(c) Parameters ${\cal C}_{1}$ (dashed lines) and ${\cal C}_{2}$ (solid). 
Different colors correspond to the different orientations of the state at time $t_{\rm rot}$, as in panel (b), while the black lines correspond to the local evolution during the preparation.
%case. 
%
The gray-shaded area is only accessible by bipartite entangled states.  
For both figures (b) and (c), we have taken $N = 1000$. 
The numerical methods used here are detailed in Supplemental Material Section G.
}
\label{Fig2}
\end{figure}
%%%%%%%%%%%%%%%%%%%%%%%%%%%%%%%%%%%%%%
%%%%%%%%%%%%%%%%%%%%%%%%%%%%%%%%%%%%%%
%%%%%%%%%%%%%%%%%%%%%%%%%%%%%%%%%%%%%%

{\bf Many-body GID.}
Our second protocol probes local decoherence induced by gravitational nonlocal dynamics.
As in the previous protocol, we start from $z$-polarized ensembles and, at $t=0$, apply a $\pi/2$ pulse to prepare
$\ket{\mathrm{CSS}_x}_j$ in each BEC ($j=A,B$).
We then let the system evolve under Eq.~(\ref{eq:HGbec}) for a (dimensionless) preparation time
$\tau_{\rm rot}\equiv (\chi_{\rm loc}^G+\chi_{\rm cont})\, t_{\rm rot}$,
assuming $\chi_{\rm loc}^G+\chi_{\rm cont}\gg \chi_{\rm nloc}^G$ so that the inter-node coupling
$\hat J_A^{z}\otimes \hat J_B^{z}$ is negligible during this stage.
This generates local spin squeezing via one-axis twisting~\cite{KitagawaPRA1993}.
Denoting the resulting local states by $\ket{\psi(\tau_{\rm rot})}_j$, we apply fast local rotations by an angle $\theta$
(about the $x$ axis for definiteness),
$\ket{\psi_j}=e^{-i\theta \hat J_j^x}\ket{\psi(\tau_{\rm rot})}_j$,
which we assume to be effectively instantaneous on the timescale of the Hamiltonian dynamics.
Finally, we switch off the local evolution by tuning $|\chi_{\rm loc}^G+\chi_{\rm cont}|\ll \chi_{\rm nloc}^G$
and let the system evolve freely under the nonlocal coupling $\chi_{\rm nloc}^G\,\hat J_A^{z}\otimes \hat J_B^{z}$.
The full sequence is shown schematically in Fig.~\ref{Fig2}(a).

Although the global evolution is unitary and entangling, its reduced action on subsystem $A$ appears as a conditional rotation.
Averaging over the $J_B^z$ distribution yields
\be \label{rhoA}
\rho_A(\tau)=\sum_{\mu_B} p(\mu_B)\,
e^{-i \mu_B \tau \hat J_A^z}\ket{\psi_A}\bra{\psi_A}\,e^{+i \mu_B \tau \hat J_A^z},
\ee
where $\mu_j$ and $\ket{\mu_j}$ are the eigenvalues and eigenstates of $\hat J_j^z$, 
$p(\mu_B)=|\braket{\mu_B|\psi_B}|^2$ and $\tau = \chi_{\rm nloc}^G(t - t_{\rm rot})$.
The finite width of $p(\mu_B)$ induces a dephasing (phase-diffusion) channel on $A$:
subsystem $B$ acts as an uncontrolled degree of freedom, and tracing it out suppresses coherences in $A$.
This effect can be seen by monitoring the degradation of local spin squeezing.

A key control knob is the squeezing orientation.
Let $\theta_0$ be the rotation that aligns the locally squeezed quadrature at $\tau_{\rm rot}$; we define
$\beta\equiv \theta-\theta_0-\pi/2$, so that $\beta=0$ corresponds to squeezing along $z$.
For $\beta=0$ the $\hat J_A^z$ distribution is narrow and the subsequent dephasing is slow.
Conversely, aligning the squeezed quadrature along $y$ ($\beta=\pi/2$) makes the variance of $\hat J^z$ large and accelerates local dephasing.
This rotation-dependent dephasing rate is a characteristic signature of the nonlocal interaction in Eq.~(\ref{eq:HGbec}) with local terms switched off, as shown in Fig.~\ref{Fig2}(b).
Using the time at which $\xi_{\rm loc}^2$ returns to $1$ as an operational dephasing time, we find
$\tau_{\rm deph}\sim N^{-1.2}$ for $\beta=\pi/2$: a very favorable scaling with $N$~\cite{note3}.

In a realistic experiment, the loss of local squeezing alone does not certify that the underlying dynamics remains coherent, as in GID.
We therefore also monitor the entanglement-witness criteria in Eq.~(\ref{ent_crit}).
Figure~\ref{Fig2}(c) reports ${\cal C}_1$ and ${\cal C}_2$: observing ${\cal C}_1<1$ or ${\cal C}_2<1$ together with the decay of local squeezing provides strong evidence that the overall evolution remains coherent and genuinely nonlocal.
The criterion ${\cal C}_2$ is particularly effective for small but nonzero $\beta$ (blue dashed line), where collective squeezing continues to build up after the local terms are switched off and the nonlocal interaction is turned on.
For the most sensitive setting $\beta=\pi/2$, phase diffusion is so rapid that ${\cal C}_2$ no longer detects entanglement, whereas ${\cal C}_1$ still does, confirming that global coherence is preserved.
Finally, to observe entanglement via ${\cal C}_2$ one must choose $\tau_{\rm rot}$ such that the locally squeezed state at $\tau_{\rm rot}$ remains close to the Fisher information saturating regime (black dashed line); otherwise ${\cal C}_2$ stays above unity even if additional collective squeezing is generated (e.g.\ for $\beta=\pi/24$).

%%%%%%%%%%%%%%%%%%%%%%%%%%%%%%%%%%%%%%
%% FIGURE 4
%%%%%%%%%%%%%%%%%%%%%%%%%%%%%%%%%%%%%%
\begin{figure}[t!]
\includegraphics[width=0.42\textwidth]{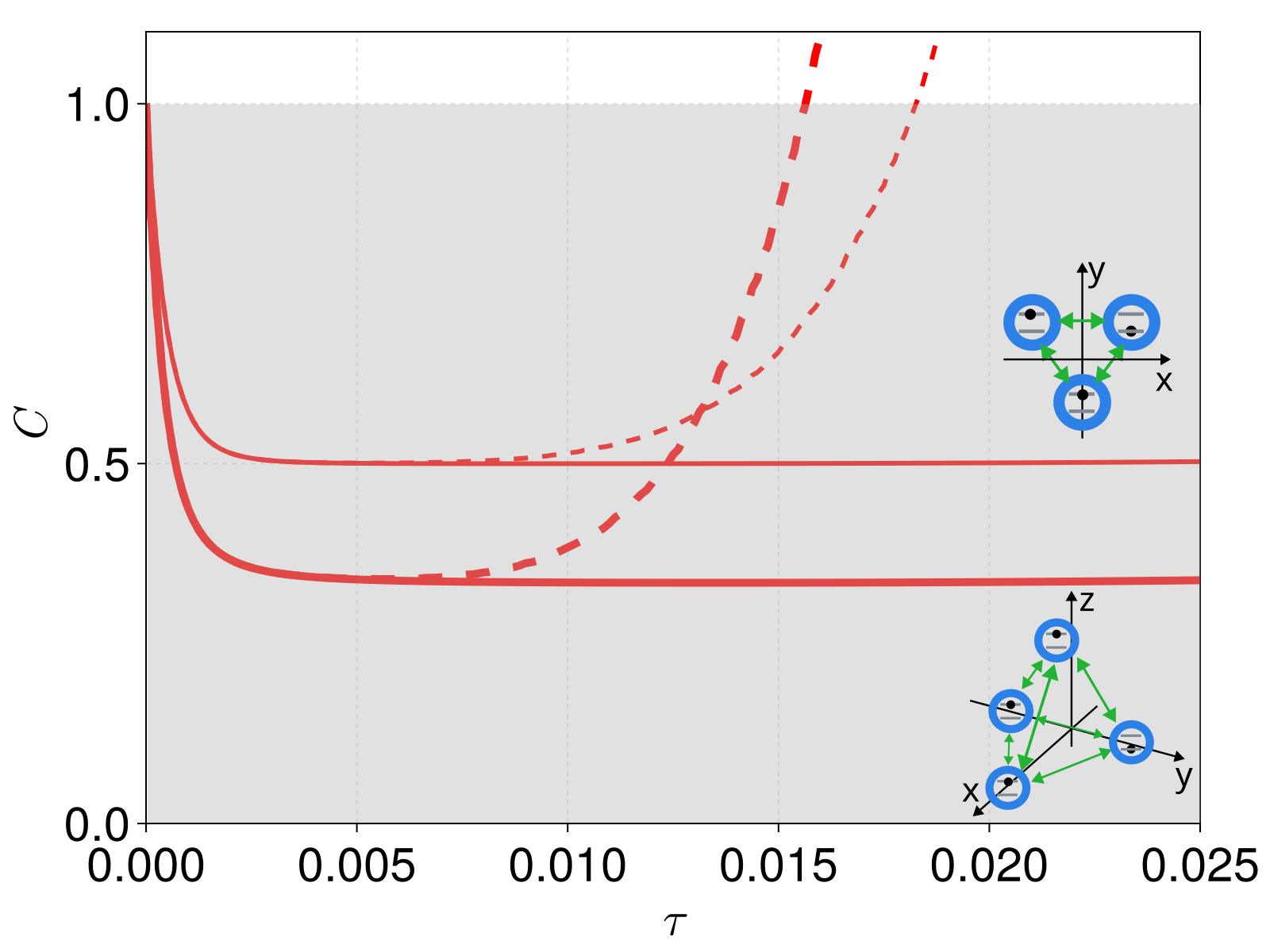}
\caption{{\bf GIE dynamics in a network of 3 and 4 nodes.} 
Quantities ${\cal C}_{1}$ (red solid) and ${\cal C}_{2}$ (red dashed) as a function of $\tau = \chi_{\rm nloc} t$.
The protocol generalizes the one described in Fig ~\ref{Fig1} (a), with the initial state at time $\tau=0$ being a product of coherent spin states $\bigotimes_{j=1}^M \ket{CSS_x}_j$, for which ${\cal C}_{1}={\cal C}_{2}=1$.
Thin lines are for $M=3$ BECs, while thick lines are for $M=4$ BECs. 
The gray region witnesses entanglement according to ${\cal C}_{1}, {\cal C}_{2}<1$.
Insert: schematic of the network with 3 and 4 nodes. Each BEC is represented by a circle with two modes.
In the numerics we have used $N= 1000$ in each ensemble. 
All the analytical formulae used to obtain these results are detailed in Supplemental Material Section H.}
\label{Fig3}
\end{figure}
%%%%%%%%%%%%%%%%%%%%%%%%%%%%%%%%%%%%%%
%%%%%%%%%%%%%%%%%%%%%%%%%%%%%%%%%%%%%%
%%%%%%%%%%%%%%%%%%%%%%%%%%%%%%%%%%%%%%

{\bf Many-body GIE and GID in quantum networks.}
The ideas above extend naturally to a network of $M$ ensembles, or bimodal interferometers.
In fact, assuming uniform couplings, the Hamiltonian in Eq.~\eqref{eq:HGbec} generalizes to
\begin{equation}
\hat H_{\rm G}=
(\chi_{\rm loc}^{G}+\chi_{\rm cont})\sum_{j=1}^M (\hat J^z_{j})^2
+\chi^{G}_{\rm nloc} \sum_{\substack{i,j =1 \\ i<j}}^M \hat J^z_{i}\otimes \hat J^z_{j}.
\end{equation}
All-to-all connectivity can be realized by placing $M=3$ nodes at the vertices of an equilateral triangle (planar geometry) or $M=4$ nodes at the vertices of a regular tetrahedron (three-dimensional geometry).
A direct analogue of the gravitational $1/d$ scaling requires equal pairwise separations; for dipolar interactions ($\propto 1/d^3$), an effective renormalization to a $1/d$ form is possible only when all inter-node distances are equal~\cite{note2}.
The entanglement witnesses generalize accordingly: ${\cal C}_{1}$ becomes
${\cal C}_{1}\equiv 4M\,\Gamma_{\rm loc}/F_{\rm col}$, while ${\cal C}_{2}$ is unchanged.

Figure~\ref{Fig3} analyzes {many-body} GIE for $M=3$ and $M=4$ nodes, starting from the product state
$\bigotimes_{j=1}^M \ket{\mathrm{CSS}_x}_j$.
As in the two-node case, we observe that criterion ${\cal C}_{1}<1$ remains effective at all times, whereas ${\cal C}_{2}<1$ is achieved at early times.
Notably, increasing $M$ lowers the minima of both criteria (to $\approx 0.5$ for $M=3$ and $\approx 0.4$ for $M=4$) and, more importantly, broadens the minimum region reached by ${\cal C}_{1}$ and ${\cal C}_{2}$, compared to the $M=2$ case.
This robustness against timing imperfections is a key advantage of the network architecture, facilitating an experimental observation of GIE and hence supporting the coherence of the underlying interaction.
A generalization of the GID protocol to $M$ ensembles is also possible: we obtain a behavior qualitatively similar to Fig.~\ref{Fig2}; details are provided in Supplemental Material Section H.

\section{Discussion and conclusions}

Dipolar Bose-Einstein condensates have been already identified as a robust and sensitive platform to probe the interface between quantum mechanics and gravity within the framework of gravitational quantum state reduction~\cite{Howl2019}, where quantum mechanics is modified by introducing a gravity-induced collapse mechanism. Here, we instead preserve the linear structure of quantum mechanics and propose a mapping of the CGB and BMV scenarios onto bimodal BECs. This mapping allows for a common description of these paradigmatic experiments that: i) is consistent with the mass-energy equivalence principle within the framework of linearized quantum gravity~\cite{Anastopoulos2014, Chen2025}; ii) substantially increases the visibility of GIE and GID relative to the original single-qubit proposals~\cite{Bose2017, Marletto2017, Castro-RuizPNAS2017}; and iii) naturally provides the implementation of an analogue platform~\cite{PetruccianiNATCOMM2026} for gravitating quantum systems, to evaluate the detectability and metrological bounds of the proposed signatures.

Regarding point (i): our derivations of the BECs mapping in Eq.~\eqref{eq:HGbec} -- both in the main text and the Supplemental Material -- adopt a pedagogical and operational approach, highlighting how this naturally emerges from the interplay between the superposition principle and the mass-energy equivalence. 
Remarkably, the gravitational Hamiltonian introduced in Ref.~\cite{Chen2025} provides a more rigorous and general derivation within the framework of linearized quantum gravity; our result is recovered by implementing Dirac quantization in the energy eigenbasis and integrating out the spatial degrees of freedom. 
As a consequence, any configuration within the BMV and CGB setups associated with different energy eigenvalues represents a physically distinct state~\cite{diffeo}.
Ultimately, both frameworks are designed to test whether the gravitational field can exist in a quantum superposition. Building on this principle, we implement such a test by exploiting the interplay between local and collective squeezing.

Regarding point (ii): while previous studies have focused on the gravitational interaction responsible for the accumulated phase, the role of gravitational self-energy in the entangling dynamics has traditionally been neglected or dismissed as a trivial mass renormalization~\cite{Chen2025, Anastopoulos2014, Castro-RuizPNAS2017}. In contrast, this many-body description reveals that the self-energy term manifests as local spin-squeezing, which can alter and enhance the entanglement generation. 
Although using effective qudits is not entirely novel, our approach introduces distinct advantages by leveraging these squeezing terms during the preparation procedure, alongside the unique physical properties of BECs. 
Specifically, existing generalizations of the BMV setup to large spins~\cite{Braccini2023} rely on single-particle, multi-arm interferometers, showing a significant increase in entanglement entropy and a corresponding drop in negativity when the number of arms $N$ reaches the tens. 
%On the CGB side, Fromonteil et al.~\cite{Fromonteil2025} proposed using an effective qudit atomic ensemble to generate non-local mass superpositions (NOON-like states) in a network. Compared to the latter, our framework provides robust access to the entire qudit structure while achieving a decoherence time that scales optimally with $N$. 
Compared to that, our many-body description uncovers a previously unrecognized role of the squeezing parameters in the entanglement witness, which is significantly enhanced through the interplay between local and collective squeezing. 

Another advantage of the proposed schemes is that they do not require using Heisenberg-scaling states as NOON states, which are very fragile to atomic losses. Instead, squeezed states are enough to see a strong improvement in the scaling of $t_{\rm min}$ and $t_{\rm deph}$, while being more robust to experimental imperfections and particle losses \cite{LiEPJB2009}.
Furthermore, as highlighted in Ref.~\cite{Howl2019}, dipolar BECs offer a complementary and cleaner platform for testing GIE compared to concurrent solid-state proposals. BECs benefits from unparalleled experimental control, allowing interactions to be precisely tuned via Feshbach resonances~\cite{ChinRMP2010}, at the cost of weaker gravitational coupling strength. This tunability is doubly advantageous: it allows the generation of highly squeezed collective states to enhance phase sensitivity, while also permitting the interactions to be tuned to a zero-crossing to isolate the gravitational signal from electromagnetic backgrounds.
It is important to note, however, that the gains achieved through our scaling are partially offset by the optical scale of the transition energies for $\Delta E$ and the limited atom number in each condensate. 
As a result, when evaluating realistic parameters under current technologies, the expected characteristic timescale for generating detectable entanglement remains prohibitively long—on the order of the age of the universe.
Although dilute atomic superpositions of thousands of atoms do not yet represent macroscopic masses in the classic sense, their many-body nature and extreme coherence make them a vital testbed for quantum gravity protocols.
Our model thus establishes a rigorous theoretical foundation that can be fully exploited as future quantum technology upgrades become available, e.g. in the atom number, which would fully benefit from the progress achieved by the proposed protocols.  
%as with the current capabilities the time needed to see GIE or GID effects remains largely out-of-reach despite the progress achieved by the proposed protocols.}

Regarding point iii): the common description in terms of bimodal BEC naturally calls for an analogue realization with stronger long-range, e.g. dipolar, interactions.
%, as with the current capabilities the time needed to see GIE or GID effects remains largely out-of-reach despite the progress achieved by the proposed protocols.
In fact, spatially-separated bimodal BECs with long-range interactions realize an effective nonlocal coupling that is formally equivalent to the gravitational Hamiltonian underlying both distant composite gravitating quantum clocks (CGB-type) and many-body gravitating interferometers (BMV-type). Moreover, the experimental platform realized in Ref.~\cite{PetruccianiNATCOMM2026} is particularly promising for realizing such analogue experiments.
There, an array of double wells is used to trap a potassium BEC with dipolar interactions and tunable scattering length: coherence times on the order of $1~\mathrm{s}$, guaranteed by the trapping system, suggest that dipolar-induced GIE/GID-like dynamics should be observable.
Using the parameters of Ref.~\cite{PetruccianiNATCOMM2026}, we estimate $\chi_{\rm nloc}\approx 4.5\times 10^{-4}\,\mathrm{s}^{-1}$.
For these values, the GIE witness ${\cal C}_2$ reaches its minimum at
$t_{\min}\sim 3.2~(0.8)\,\mathrm{s}$ for $N=1000~(4000)$ atoms [cf.\ Fig.~\ref{Fig1}(b)].
For the GID protocol, the operational dephasing time at $\beta=\pi/2$ [red curve in Fig.~\ref{Fig2}(b)] is
$t_{\rm deph}\sim 0.78~(0.17)\,\mathrm{s}$ for $N=1000~(4000)$.
These times are comparable with the coherence time of current BEC platforms, which can be increased with spin echo techniques \cite{EgorovPRA2011} to preserve the visibility of the GIE and GID mechanisms.
This platform would hence allow the realization of a programmable analogue gravitating quantum dynamics where it is possible to explore regimes, entanglement witnesses, and dephasing scalings that are otherwise currently inaccessible in genuine gravity experiments.

In conclusion, our work opens toward optimizing the probe states to further enhance the visibility of GIE and GID, and to exploiting controlled local nonlinearities during the entangling stage to amplify the signatures of nonlocal dynamics. 
Furthermore, access to a highly sensible analogue platform paves the way for investigating the physics of quantum reference frames -- such as the temporal localizability of events when employing interacting quantum clocks~\cite{Castro-Ruiz2020}, and the implications of a generalized (quantum time-dilated) Schrödinger equation~\cite{SmithNATCOMM2020, CafassoPRD2024}. 
Beyond the frameworks preserving standard quantum theory, other approaches to quantum gravity may also benefit from these ideas. Specifically, tracing out the environment -- comprising one of the BECs and the gravitational field -- enables analogue simulation and benchmarking of gravitational quantum state reduction~\cite{Howl2019} or classical diffusion models~\cite{AngeliARXIV}.
Finally, beyond the scope of fundamental physics, the analog platform, and the proposed protocols, would serve as a versatile toolbox for quantum sensing, working as a highly sensitive apparatus to detect weak distance-dependent long-range interactions between spatially separated BECs.
 
\section{Methods}

{\bf Qudit spin operators from bosonic modes.}
We consider two pairs of bosonic modes with bosonic annihilation (creation) operators given by $\hat{c}_{\uparrow j}$  and $\hat{c}_{\downarrow j}$ ($\hat{c}_{\uparrow j}^\dag$  and $\hat{c}_{\downarrow j}^\dag$) with $j=A,B$ labeling the subsystem. Using a Schwinger transformation, we can build spin operators out of this modes, defined by:
\begin{align}
    \hat{J}^x_{j} &= [\hat{c}_{\uparrow j}^\dagger \hat{c}_{\downarrow j} + \hat{c}_{\downarrow j}^\dagger \hat{c}_{\uparrow j}]/2 \\
    \hat{J}^y_{j} &= [\hat{c}_{\uparrow j}^\dagger \hat{c}_{\downarrow j} - \hat{c}_{\downarrow j}^\dagger \hat{c}_{\uparrow j}]/2i \\
    \hat{J}^z_{j} &= [\hat{c}_{\uparrow j}^\dagger \hat{c}_{\uparrow j} - \hat{c}_{\downarrow j}^\dagger \hat{c}_{\downarrow j}]/2
\end{align}
These are pseudo-spin operators for effective qudits of length $S=N/2$ satisfying commutation relations analogous to those of collective spin operators, namely $[\hat{J}^l_{j}, \hat{J}^m_{i}] = i \epsilon_{l,m,k} \hat{J}^k_j \delta_{i,j}$ with $\epsilon_{l,m,k}$ being the Levi-Civita symbol. In the main text, we considered the case where $\uparrow$ and $\downarrow$ correspond to internal degrees of freedom (quantum clock) or external degrees of freedom (interferometer arm or external trapping potential).  \\

{\bf Quantum Hamiltonian for two gravitating qudits.}
Following the same reasoning as CGB but for two $N+1$-level qudits, placed at distance $d$ and with even energy-level spacing $\Delta E$, we obtain the following Hamiltonian:
\begin{equation}\label{eq:HGint}
    %\hat{H}_{G}^{qudit} = \chi_G \,\sigma_A^z \otimes \sigma_B^z \,,
    \hat{H}_{\rm G}^{\rm qudit} = \chi^G_{nloc}~\hat{J}^z_A \otimes \hat{J}^z_B 
    \,,
\end{equation}
with $\chi^{G}_{\rm nloc} = G \Delta E^{2}/d c^{4}$. The derivation of Eq.~\eqref{eq:HGint}, which generalizes the CGB proposal to two qudits, is provided by both adopting a fully quantum (Supplemental Material Section~D) and a quantum field theory (Supplemental Material Section~E) approach. We emphasize that this Hamiltonian is equivalent to the N-particle BEC Hamiltonian provided one can cancel the local non linear terms due to BEC self interaction. This Hamiltonian can also be generalized to the case of a network of $M$ equidistant qudits:
\begin{equation}\label{eq:HGintM}
    %\hat{H}_{G}^{qudit} = \chi_G \,\sigma_A^z \otimes \sigma_B^z \,,
    \hat{H}_{\rm net}^{\rm qudit} = \chi^G_{\rm nloc}~ \sum_{1\leq i< j \leq M} \hat{J}^z_i \otimes \hat{J}^z_j 
    \,,
\end{equation}
which is again equivalent to the network BEC Hamiltonian provided we cancel local terms.

{\bf Multiparameter squeezing and Fisher information for entanglement detection.} \\
In the context of multi-parameter estimation, the squeezing parameter can be generalized to an ensemble of $M$ subsystems via the squeezing matrix \cite{gessner_multiparameter_2020}. It is an $M \times M$ symmetric matrix ($M$ being the number of subsystems) $\vect{\xi}^{2}$ where the diagonal elements are the local Wineland squeezing parameter~\cite{winelandPRA1994},
\begin{equation}
    \label{sqz_loc}
    \xi_{jj}^{2} \left( \theta_{j} \right) = \frac{ N_{j}  \mathrm{Var} \left( c_{j}  J^{y}_{j} - s_{j}  J^{z}_{j}  \right)}{\vert \langle J^{x}_{j} \rangle \vert ^{2}}
\end{equation}
while the off-diagonal elements are
\begin{align} \label{xi12}
\xi_{ij}^{2} \left( \theta_{i}, \theta_{j} \right) %= \xi^{2}_{ji} \left( \theta_{i}, \theta_{j} \right) \nonumber \\
=  \frac{\sqrt{ N_{i} N_{j} } \langle \left[ c_{i} J^{y}_{i} - s_{i} J^{z}_{i} \right]  \left[ c_{j}  J^{y}_{j}  - s_{j} J^{z}_{j} \right]\rangle }{\vert \langle  J^{x}_{i} \rangle \langle J^{x}_{i} \rangle \vert}
\end{align}
with $c_{j} = \cos \theta_{j}$ and $s_{j} = \sin \theta_{j}$, with $i,j = 1...M$ labeling the different subsystems. 
The collective squeezing is then obtained by
computing the expectation value $\vect{\nu}^\top\vect{\xi} \vect{\nu}$ with respect to arbitrary $M$-dimensional real vector $\vect{\nu}$ and by optimizing with respect to the orientation angles $\theta_j$: $\xi^{2}_{\rm col} = min_{\vect{\theta}} \vect{\nu}^{\top} \vect{\xi}^{2} (\vect{\theta}) \vect{\nu}$.
In the main text, we only considered the uniform case $\xi^{2}_{\rm col} = \xi^{2}_{\rm unif}$ corresponding to $\vect{\nu} = (1 ... 1)^{T}/\sqrt{M}$ for the collective squeezing. 
On the other hand, the local spin squeezing is simply given by: $\xi^{2}_{\rm loc} = min_{\theta_{1}} \xi_{11}$ as we assume that we have a permutation symmetry between the different subsystems.
The difference between the local and collective squeezing can serve as an entanglement criterion to detect entanglement between the subsystem, however it works only for pure states. For the most general case with mixed states, we use the criteria given in Eq.~\eqref{ent_crit}, based on the covariance matrix $\Gamma$ and the Fisher matrix $\vect{F}$ as defined in Ref.~\cite{LiPRA2013, GessnerPRA2016}. The covariance matrix is a $3M \times 3M$ matrix, whose elements are given by:
\begin{equation} 
    \Gamma_{3i+m, 3j+n} = \langle J_{i}^{\alpha_{m}} J_{j}^{\alpha_{n}} \rangle -  \langle J_{i}^{\alpha_{m}} \rangle \langle J_{j}^{\alpha_{n}} \rangle
\end{equation}
with $i, j \in \left[1...M\right]$, $m, n \in (1, 2, 3)$ and $\alpha_{1} =x, \alpha_{2}=y, \alpha_{3}=z$.
$\Gamma_{\rm col}$ is the largest eigenvalue of the covariance matrix $\vect{\Gamma}$ while $\Gamma_{\rm loc}$ is the largest eigenvalue of the $3 \times 3$ upper left block of the covariance matrix - corresponding to the covariance matrix of observables for the subsystem 1.
The Fisher matrix is also a $3M \times 3M$ matrix, and requires the diagonalization of the density matrix. If we assume that the density matrix can be written as $\rho = \sum_{k} p_{k} \vert \Psi_{k} \rangle \langle \Psi_{k} \vert$ (with $\sum_{k} p_{k} = 1, ~p_{k} \geq 0 \forall k$), then the elements of the Fisher matrix take the following form:
\begin{equation} 
    F_{3i+m, 3j+n} = 2 \sum_{\substack{k, l\\ p_{k}+p_{l} > 0}}  \frac{(p_{k} - p_{l})^{2}}{p_{k} + p_{l}} \langle \Psi_{k} \vert J_{i}^{\alpha_{m}} \vert \Psi_{l} \rangle \langle \Psi_{l} \vert J_{j}^{\alpha_{n}} \vert \Psi_{k} \rangle.
\end{equation}
Then, $F_{\rm col}$ is the largest eigenvalue of the $3M \times 3M$ Fisher matrix $\vect{F}$ of the total system, while $F_{\rm loc}$ is the largest eigenvalue of the $3 \times 3$ Fisher matrix obtained from the reduced density matrix on subsystem 1, with $\rho_{1} = {\rm Tr}_{2...M} ( \rho)$.
If the total system is pure, we always have $F_{col} = 4 \Gamma_{col}$. However, this is not true for the local quantities, and in the presence of entanglement, one can have $F_{loc} < 4 \Gamma_{loc}$. From the local Fisher information $F_{\rm loc}$, it is possible to derive tighter bounds than Eq. (\ref{ent_crit}):
\begin{equation}
    \tilde{\cal C}_{1} \equiv \frac{M F_{\rm loc}}{F_{\rm col}}, \quad \tilde{\cal C}_{2} \equiv \frac{\xi^{2}_{col} F_{\rm loc}}{N}
\end{equation}
such that we have $\tilde{\cal C}_{1} \geq 1$ and $\tilde{\cal C}_{2} \geq 1$ for all separable states. These criteria verify $\tilde{\cal C}_{1} \leq {\cal C}_{1}$  and $\tilde{\cal C}_{2} \leq {\cal C}_{2}$, hence they detect more efficiently the presence of entanglement, however they require to measure quantities that are not easily accessible experimentally, such as a full tomography of the density matrix. This is why we considered mainly criteria-\eqref{ent_crit} in the main text. 

\section{Acknowledgments}

The authors are grateful to Flaminia Giacomini for insightful inputs during the review phase. Y.T. and D.C. thank Natalia Salomè Moller for valuable feedback. D.C. also thanks Nicola Pranzini, Tomohiro Fujita, and Ivette Fuentes, for helpful discussions. 
We acknowledge support from the QuantERA project SQUEIS (Squeezing enhanced inertial sensing), funded by the European Union's Horizon Europe Program and the Agence Nationale de la Recherche (ANR-22-QUA2-0006). 
This publication has received funding under Horizon Europe programme HORIZON-CL4-2022-QUANTUM-02-SGA via the project 10113690 PASQuanS2.1; and from the PRIN 2022 Project ``Quantum Sensing and Precision Measurements with Nonclassical States''.

%%%%%%%%%%%%%%%%%%%%%%%%%%%%%%%%%%%%%%%%%%%%%%%%%%%%%%%%%%%%%%%%

\clearpage

\onecolumngrid

\begin{center}
{\bf Supplemental Material}
\end{center}

\section*{A- Interaction Hamiltonian from mass-energy equivalence for two bimodal BECs}

Consider a qubit with energy gap $\Delta E$ between its ground $\vert \downarrow \rangle$ and excited $\vert \uparrow \rangle $ states and resting mass $m$. According to the core assumption of CGB~\cite{Castro-RuizPNAS2017}, the gravitational mass can be replaced by an operator as $ m \longrightarrow \hat{m} \equiv m + \frac{\Delta E}{c^{2}} \sigma^{z}$.
In this work, we consider two composite quantum clocks~\cite{CafassoPRD2024}, denoted by $j= A, ~B$, each made up of $N$ particles in a bimodal BEC. The atoms in each BEC can thus occupy two modes, denoted by $\downarrow$ and $\uparrow$, with energy gap $\Delta E$. Furthermore, a spatial wavefunction $\psi_{\sigma j} ( \mathbf{r})$, with $\sigma = ~\downarrow, ~\uparrow$, as well as a local creation (resp. annihilation) operator $\hat{c}^{\dagger}_{\sigma j} $ (resp $\hat{c}_{\sigma j}$), is associated to each mode. The interaction Hamiltonian is then given by
\be
\hat{H}_{\rm CGB} = \frac{1}{2} \sum_{\sigma_{1}, \sigma_{2}, \sigma_{3}, \sigma_{4}} \int d\mathbf{r} \int d\mathbf{r}'~ \Psi_{\sigma_{1}}^{\dag}(\mathbf{r}) \Psi_{\sigma_{2}}^{\dag}(\mathbf{r}') U(\mathbf{r}-\mathbf{r}') \Psi_{\sigma_{3}}(\mathbf{r})\Psi_{\sigma_{4}}(\mathbf{r}')\,,
\ee
where the field operator
\begin{equation}
\hat{\Psi}_{\sigma}(\mathbf{r}) =  \psi_{\rm \sigma A}(\mathbf{r}) \hat{c}_{\rm \sigma A} + \psi_{\rm \sigma B}(\mathbf{r}) \hat{c}_{\rm \sigma B}  \,.
\end{equation}
In this description, we consider contact interaction within the same BEC, and long-range gravitational interaction coupling both atoms within the same BEC and the two quantum clocks. Since these interactions do not change the internal states during the evolution (i.e. $U$ is diagonal in this basis), we always have $\sigma_{1} = \sigma_{3}$ and $\sigma_{2} = \sigma_{4}$. Thus, we can rewrite the interaction Hamiltonian as
\be
\hat{H}_{\rm CGB} = \frac{1}{2} \sum_{\sigma, \sigma'} \int d\mathbf{r} \int d\mathbf{r}'~ \Psi_{\sigma}^{\dag}(\mathbf{r}) \Psi_{\sigma'}^{\dag}(\mathbf{r}') U_{\sigma \sigma'}(\mathbf{r}-\mathbf{r}') \Psi_{\sigma}(\mathbf{r})\Psi_{\sigma' }(\mathbf{r}')\,,
\ee
with
\be \label{Upot}
U_{\sigma \sigma'}(\mathbf{r}-\mathbf{r}') 
= g_{\sigma \sigma'}\delta(\mathbf{r}-\mathbf{r}') 
+ \frac{C_{\sigma \sigma'} }{\vert \mathbf{r} - \mathbf{r}'\vert}\,,
\ee
where $g_{\sigma \sigma'} \approx 4 \pi \hbar^2 a_{\sigma \sigma'}/m$ is the contact interaction coupling, with $a_{\sigma \sigma'}$ the scattering length associated with different internal energy levels, and $C_{\sigma \sigma'} = G m_{\sigma} m_{\sigma'}$ is the gravitational interaction coupling, with $ m_{\downarrow} = m, ~m_{\uparrow} = m + \Delta E / c^{2}$, and
$G = 6.67 \times 10^{-11}~{\rm m }^{3}~{\rm kg}^{-1} ~{\rm s}^{-2}$ Newton's gravitational constant. 

In what follows, we assume that the interaction does not change the mode profile, i.e. the spatial wavefunction, but only generate dynamics at the level of the mode occupations. In addition, we also assume that $\psi_{\uparrow A}$ and $\psi_{\downarrow A}$ have the same spatial profile, and the wavefunctions for $B$ can be obtained simply by a translation. Then, introducing the spin operators 
\begin{equation}
\hat{J}^{z}_{j} = \frac{\hat{c}_{\rm \uparrow j}^\dag \hat{c}_{\rm \uparrow j} - \hat{c}_{\rm  \downarrow j}^\dag\hat{c}_{\rm \downarrow j}}{2}\,,
\end{equation}
we can map each BEC onto a $N+1$-level qudit, with effective spin length $S = N/2$. Hence, it is possible to rewrite the interaction Hamiltonian in the following way:
\begin{align}
\label{ham_int1}
\hat{H}_{\rm CGB} &= \frac{g_{\uparrow \uparrow} + g_{\downarrow \downarrow} - 2 g_{\uparrow \downarrow}}{2} I \Bigg( \left( \hat{J}^{z}_{A} \right)^{2} + \left( \hat{J}^{z}_{B} \right)^{2} \Bigg) +  \frac{g_{\uparrow \uparrow} - g_{\downarrow \downarrow}}{2} I \Bigg( \left(N-1\right) \hat{J}^{z}_{A}  + \left(N-1\right)\hat{J}^{z}_{B} \Bigg)  \nonumber \\
&+ \frac{C_{\uparrow \uparrow} + C_{\downarrow \downarrow} - 2 C_{\uparrow \downarrow}}{2}  \Bigg( D_{AA} \left( \hat{J}^{z}_{A} \right)^{2} + D_{BB} \left( \hat{J}^{z}_{B} \right)^{2} \Bigg) +  \frac{C_{\uparrow \uparrow} - C_{\downarrow \downarrow}}{2} \Bigg( D_{AA} \left(N-1\right) \hat{J}^{z}_{A}  + D_{BB} \left(N-1\right)\hat{J}^{z}_{B} \Bigg)  \nonumber \\  
&+ D_{AB} \left[ \left( C_{\uparrow \uparrow} + C_{\downarrow \downarrow} - 2 C_{\uparrow \downarrow} \right) \hat{J}^{z}_{A}  \hat{J}^{z}_{B} +  \left( C_{\uparrow \uparrow} - C_{\downarrow \downarrow} \right) \left( N \hat{J}^{z}_{A}  + N \hat{J}^{z}_{B} \right) \right]
\end{align}
where we have omitted constant terms, and introduced the spatial integrals
\begin{equation}
    I\equiv \frac{1}{2}\int d\mathbf{r} ~ \vert \psi_{\sigma j}(\mathbf{r}) \vert ^{2} \vert \psi_{\sigma' j}(\mathbf{r}) \vert ^{2} \,,
\end{equation}
for contact interactions, and 
\begin{equation}\label{eq:Dij}
    D_{ij} \equiv  \frac{1}{2}\int d\mathbf{r}~  \vert \psi_{\sigma i}(\mathbf{r}) \vert ^{2} \int d\mathbf{r}'~   \frac{ \vert \psi_{\sigma' j}(\mathbf{r}' ) \vert^{2}}{\vert \mathbf{r} - \mathbf{r}'\vert~}\,,
\end{equation}
for the gravitational interaction within ($i=j)$ and between ($i\neq j$) the two BECs. 
Notice that no further regularization is needed in Eq.~\eqref{eq:Dij} for $i=j$ since we assume that the wavefunctions have a finite support, and no $\sigma$-dependence is left in $I$ and $D_{ij}$ - given that each mode has the same spatial profile, which is left within $g_{\sigma \sigma'}$ for contact and $C_{\sigma \sigma'}$ for gravitational interaction. 
Thus, $I$ does not depend on the internal level nor the subsystem, while $D_{ij}$ only depends on the choice of subsystems (if both $i$ and $j$ are equal or not).
Furthermore, in order to obtain the Hamiltonian in Eq.~\eqref{ham_int1}, we have neglected all the integrals that involve the overlap of two different wavefunctions $\psi_{i \sigma }^{\dagger} \left( \mathbf{r} \right)  \psi_{j \sigma} \left( \mathbf{r} \right) $ ($i \neq j$), as the wavefunctions do not overlap in real space. Finally, for symmetry reason, we have that $D_{AA} = D_{BB} \equiv D$, and if the distance $d$ is large compared to the typical size of the BEC, we also have $D_{AB} \approx 1/d$, leading to:
\begin{align}
\hat{H}_{\rm tot} &= \chi_{Nz}  \left( \hat{J}^{z}_{A} + \hat{J}^{z}_{B} \right) + \chi_{\rm G} \hat{J}^{z}_{A} \hat{J}^{z}_{B} + (\chi_{\rm loc}+\chi_{\rm cont}) \left( (\hat{J}^{z}_{A})^{2} + (\hat{J}^{z}_{B})^{2} \right) 
\end{align}
where:
\begin{align}
\chi_{\rm cont} &= I\, (g_{\uparrow \uparrow} + g_{\downarrow \downarrow} - 2 g_{\uparrow \downarrow}) \\
\chi_{\rm loc}^{G} &= D \frac{ G(\Delta E)^{2}}{c^{4}}\\
\chi_{\rm nloc}^{G} &= \frac{1}{d}\frac{G(\Delta E)^{2}}{c^{4}}\\
\chi_{Nz} &= \left[ I (g_{\uparrow \uparrow} - g_{\downarrow \downarrow})+ D\frac{2 Gm \Delta E}{c^{2}}  \right](N-1) + \frac{2 G m \Delta E}{dc^{2}} N  .
\end{align}
where we used $C_{\uparrow \uparrow} - C_{\downarrow \downarrow} \approx 2 m \Delta E/c^{2}$, assuming $m \gg \Delta E / c^{2}$ and taking $\hbar = 1$.

The two local linear terms can be compensated by a spin-echo protocol, leading to the Hamiltonian given in Eq. (1) in the main text. Moreover, the value of $\chi_{\rm cont}$ can be tuned using a Feshbach resonance. With this, it is possible to completely cancel the local terms, and we obtain exactly the qudit Hamiltonian given in Eq. (8) in the main text. Note that the value obtained for $\chi_{G}$ here is exactly the same as the one obtained from the qudit derivation, see Appendix D.

\section*{B- Interaction Hamiltonian via non-local mass superposition for two bimodal BECs}

Consider two quantum systems, denoted $A$ and $B$, traveling through a matter-wave interferometer in the BMV setup~\cite{Bose2017,Marletto2017}. The total Hamiltonian in the weak-gravity regime depends on the position operators as
\begin{equation}
    H_{\rm BMV} = m_A c^2+ m_B c^2- \frac{G \,m_Am_B}{|\hat x_A - \hat x_B|} \,,
\end{equation}
where the interaction term is expected to give rise to a phase shift, eventually witnessing the non-classical nature of gravity. 

The spatial superposition of the masses gives rise to the four-level spectrum
\begin{equation}
    \left\{
    \begin{aligned}
    E_{\downarrow \downarrow} &= m_A c^2+ m_B c^2 - \frac{G \,m_A m_B}{d+2d'}
    \\
    E_{\downarrow \uparrow} &= m_A c^2+ m_B c^2 - \frac{G \,m_A m_B}{d+d'}
    \\
    E_{\uparrow \downarrow} &= m_A c^2+ m_B c^2 - \frac{G \,m_A m_B}{d+d'}
    \\
    E_{\uparrow \uparrow} &= m_A c^2+ m_B c^2- \frac{G \,m_A m_B}{d}
    \end{aligned}
    \right.\,,
\end{equation}
where 
the arrows denote the interior $\uparrow$ and exterior $\downarrow$ arms of the interferometers for $A$ and $B$ respectively, and
we introduced the eigenvalue of the relative-distance operator $|\hat x_A - \hat x_B|$ as $d$ when $A$ and $B$ are the closest, and $d'$ is the interarm distance within a same interferometer.

Consider now the case in which, instead of a single particle in each interferometer, we send $N_{A}$ bosonic particles in $A$ and $N_{B}$ particles in $B$. The state of the total system is then given by a superposition of all the possible Fock states $\vert N_{A \downarrow}, ~N_{A \uparrow}, ~N_{B \downarrow}, ~N_{B \uparrow}\rangle$. 
Unlike the single qubit case, we now need to take into account the gravitational and contact interactions within and between each arm of a single interferometer. Introducing the spatial wavefunction for the atoms is each arm as $\psi_{\downarrow, A}, \psi_{\uparrow, A}, \psi_{\downarrow, B}$ and $\psi_{\uparrow, B}$, and assuming that the spatial profile is the same for each arm (up to a translation), the following contact interaction leads to the following phase accumulation over a time $t$:
\begin{equation}
    \phi_{\rm cont}  (N_{A \downarrow}, ~N_{A \uparrow}, ~N_{B \downarrow}, ~N_{B \uparrow})/t =  \frac{I}{2} \left[ N_{A \downarrow}(N_{A \downarrow}-1) + N_{A \uparrow}(N_{A \uparrow}-1) + N_{B \downarrow}(N_{B \downarrow}-1) + N_{B \uparrow}(N_{B \uparrow}-1) \right]
\end{equation}
with $I = g \int {\rm d} {\bf r} \vert \psi_{\sigma j} ({\bf r}) \vert ^{4}$ which is the same for all four arms. Here we have taken contact interaction in the form $g \delta (r-r')$ as we did in Appendix A, with $g = 4 \pi \hbar^2 a/m$ and $a$ the scattering length. 

Further assuming that the distance between the arms is large compared with the typical width of the spatial wavefunctions, we obtain the following phase for gravitational interaction within a single interferometers:
\begin{align}
    \phi_{\rm loc}  (N_{A \downarrow}, ~N_{A \uparrow}, ~N_{B \downarrow}, ~N_{B \uparrow})/t &= - \left[ N_{A \downarrow}(N_{A \downarrow}-1) + N_{A \uparrow}(N_{A \uparrow}-1) + N_{B \downarrow}(N_{B \downarrow}-1) + N_{B \uparrow}(N_{B \uparrow}-1) \right] D \nonumber \\
    &-  N_{A \downarrow} N_{A \uparrow} \frac{G m_{A} m_{A}}{d'} - N_{B \downarrow} N_{B \uparrow} \frac{G m_{B} m_{B}}{d'}
\end{align}
with $D = G m_{j} m_{j} \int {\rm d} {\bf r} {\rm d} {\bf r}' \vert \psi_{\sigma j} ({\bf r}) \vert ^{2} \vert \psi_{\sigma j} ({\bf r}') \vert ^{2}/(2 \vert {\bf r} - {\bf r}' \vert)$ and between the two interferometers
\begin{align}
    \phi (N_{A \downarrow}, ~N_{A \uparrow}, ~N_{B \downarrow}, ~N_{B \uparrow})_{G}/t &= N_{A \downarrow} m_A c^2 + N_{A \uparrow} m_A c^2 + N_{B \downarrow} m_B c^2 + N_{B \uparrow} m_B c^2  \nonumber \\
    &- \left[N_{A \downarrow} N_{B \downarrow} \frac{G \,m_A m_B}{d+2d'} + N_{A \downarrow} N_{B \uparrow} \frac{G \,m_A m_B}{d+d'} + N_{A \uparrow} N_{B \downarrow} \frac{G \,m_A m_B}{d+d'} +  N_{A \uparrow} N_{B \uparrow} \frac{G \,m_A m_B}{d} \right].
\end{align}

Introducing the constraints $N_{A \downarrow}+N_{A \uparrow} = N_A$, and $~N_{B \downarrow}+N_{B \uparrow} = N_B$, we can label our Fock states with only two parameters, defined as $\mu_{j} \equiv (N_{j \uparrow} - N_j/2)$, with $j=A, ~B$, such that $\mu_{j} \in [-N_{j}/2, N_{j}/2]$. From now on, we also assume that we have $m_{A} = m_{B}$. We can then simplify the expression of the phases, which, up to constant terms, reads as
\begin{align}
    \phi (\mu_{A}, ~\mu_{B})_{\rm cont}/t &=  I (\mu_{A}^{2} + \mu_{B}^{2})\nonumber \\
    \phi (\mu_{A}, ~\mu_{B})_{\rm loc}/t &=  \left(\frac{Gm^2}{d'} - D\right) (\mu_{A}^{2} + \mu_{B}^{2}) \nonumber \\
    \phi (\mu_{A}, ~\mu_{B})_{\rm nloc}/t &=  Gm^{2} \left( \frac{2}{d+d'} - \frac{1}{d} - \frac{1}{d+2d'} \right) \mu_{A} \mu_{B} + Gm^{2}\left(\frac{1}{d+2d'}-\frac{1}{d} \right) (\frac{N_{A} \mu_{B}}{2} +\frac{N_{A} \mu_{B}}{2}).
\end{align}
After a time $t$, the state $\vert \mu_{A}, ~\mu_B \rangle$ thus becomes
\begin{equation}
    \vert \mu_{A}, ~\mu_{B} \rangle (t) = e^{-i \phi (\mu_{A}, ~\mu_{B})} \vert \mu_{A}, ~\mu_B \rangle\,.
\end{equation}
This is equivalent to the unitary evolution $U = e^{-iHt}$ (assuming $\hbar = 1$) induced by the Hamiltonian
\begin{equation}
    H = (\chi_{\rm cont} + \chi_{\rm loc}^{G}) \left[ (\hat{J}^{z}_{A})^{2} + (\hat{J}^{z}_{B})^{2} \right] + \chi_{\rm nloc}^{G} \hat{J}^z_A \otimes \hat{J}^z_B + \chi_{Nz} (N_A \hat{J}^z_B + N_B \hat{J}^z_A)
\end{equation}
with
\begin{align}
    \chi_{\rm cont} &= I \\
    \chi_{\rm loc}^{G} &= \frac{Gm^{2}}{d'} - D \\
    \chi_{\rm nloc}^{G} &= -Gm^{2}\frac{2(d')^2}{d(d+d')(d+2d')} \\ 
    \chi_{Nz} &=  -Gm^{2} \frac{d'}{d(d+2d')}
\end{align}
Here, the operators $J^z_{i=A,B}$ are the spin operators for qudits of length $S_i = N_i/2$. Ignoring the linear terms (which can be compensated for with a spin-echo sequence), we finally obtain the following Hamiltonian:
\begin{equation}
    H = \left( \chi_{\rm cont} + \chi_{\rm loc}^{G} \right) [(\hat{J}^{z}_{A})^{2} + (\hat{J}^{z}_{B})^{2}] + \chi_{\rm nloc}^{G} \hat{J}^z_A \otimes \hat{J}^z_B
\end{equation}
which is exactly the same form as given in Eq.~(1) in the main text. This proves that the BMV setup can be generalized to many qubits, and that the corresponding physical description is equivalent to the case of gravitating quantum systems in a superposition of internal energy levels - derived in Appendix A.
Finally, be cancelling the local terms via a Feshbach resonance, it is possible to realize exactly the qudit Hamiltonian given in Eq.~(8) of the Methods.

Finally, taking $N_{A} = N_{B} =1$ and assuming $d' \gg d$, we recover $H= \chi_{G} \hat{\sigma}^{z}_A \otimes  \hat{\sigma}^{z}_B $ with $\chi_{G} = - Gm^{2}/d$.

\section*{C- Derivation of the Hamiltonian parameters for dipolar double-wells}

We consider a system made of two double wells, each double well acting as a subsystem. For each double-well, we can construct a linear Hamiltonian $H_{0}$ made of the sum of a kinetic term, a parabolic trap (with frequencies $\omega_x$, $\omega_y$ and $\omega_z $) and the DW potential:
\begin{equation}
    H_{0} = H_{x} + H_{y} + H_{z}
\end{equation}
In the case of a double well along $x$, we have:
\begin{align}
    H_x &= -\frac{\hbar^2}{2m} \frac{d^2}{dx^2} + V_{\rm DW}(x) +  \frac{1}{2} m \omega_x^2 x^2 \\
    H_y &= -\frac{\hbar^2}{2m} \frac{d^2}{dy^2} + \frac{1}{2} m \omega_y^2 y^2 \\
    H_z &= -\frac{\hbar^2}{2m} \frac{d^2}{dz^2} + \frac{1}{2} m \omega_z^2 z^2.
\end{align}
Here we have chosen to orient the DW potential along the $x$ direction.
The first step to diagonalize the Hamiltonian to find its ground state and first excited state, out of which we reconstruct the left-well and right-well wavefunctions. Since the three coordinates are independent, the $y$ and $z$ part of the wavefunctions can be solved analytically, and we simply obtain a Gaussian envelope for the $y$ and $z$ coordinates. However, for the $x$ coordinate, we numerically diagonalize the Hamiltonian to obtain the $\psi_{gs} \left( x \right)$ and $\psi_{ex} \left( x \right)$ wavefunctions, with energy $E_{\rm gs}$ and $E_{\rm ex}$. We can construct the left-well and right-well wavefunctions [see Fig.~\ref{Fig1} (a)] as:
\begin{equation}
    \psi_{\rm L}(x) = \frac{\psi_{\rm gs}(x) + \psi_{\rm ex}(x)}{\sqrt{2}}, ~
\psi_{\rm R}(x) = \frac{\psi_{\rm gs}(x) - \psi_{\rm ex}(x)}{\sqrt{2}}.
\end{equation}
Since the ground state and first excited state are almost degenerate, we make the two-mode approximation:
\be
\Psi(\mathbf{r}) = \psi_{\rm L}(\mathbf{r}) \hat{c}_{\rm L} + \psi_{\rm R}(\mathbf{r}) \hat{c}_{\rm R},
\ee
where $\hat{c}_{\rm L} ~(\hat{c}_{\rm R})$ is the creation operator of a particle in the left (right) well. The full expression of $\psi_{\rm L}(\mathbf{r}), ~\psi_{\rm L}(\mathbf{r})$ are given by:
\begin{equation}
\psi_{\rm L/R}(\mathbf{r}) = \psi_{\rm L/R}(x) \frac{e^{-y^2/(2 \sigma_y^2)}}{(\pi \sigma_y^2)^{1/4}} \frac{e^{-z^2/(2 \sigma_z^2)}}{(\pi \sigma_z^2)^{1/4}}
\end{equation}
where $\sigma_y^2 = \tfrac{\hbar}{m \omega_y}$ and $\sigma_z^2 = \tfrac{\hbar}{m \omega_z}$. Taking into account interactions, the total single double well Hamiltonian becomes:
\be
\hat{H} =
\hat{H}_0 + \hat{H}_{\rm int}
\ee
with
\be
\hat{H}_0 = \int d\mathbf{r}~\Psi^{\dag}(\mathbf{r}) H_0 \Psi(\mathbf{r}),
\ee
\be
\hat{H}_{\rm int} = \frac{1}{2} \int d\mathbf{r} \int d\mathbf{r}'~ \Psi^{\dag}(\mathbf{r}) \Psi^{\dag}(\mathbf{r}') U(\mathbf{r}-\mathbf{r}') \Psi(\mathbf{r})\Psi(\mathbf{r}').
\ee
Here, we consider contact interaction (within a same well) and long-range dipolar interaction coupling both atoms within the same well and between different wells:
\be \label{Upot}
U(\mathbf{r}-\mathbf{r}') = g\delta(\mathbf{r}-\mathbf{r}') + C_{dd} U_{dd}(\mathbf{r}-\mathbf{r'})
\ee
with
\be
U_{dd}(\mathbf{r}-\mathbf{r'}) =
\frac{1}{4 \pi} \frac{1-3 \cos^{2} \Theta}{\vert \mathbf{r} - \mathbf{r}'\vert^3},
\ee
where $g = 4 \pi \hbar^2 a/m$,
$\Theta$ is the angle between the $z$ axis and the vector $\mathbf{r} - \mathbf{r}'$, and $C_{dd} = \mu_0 \mu^2$ for magnetic dipoles, with 
$\mu_0 = 4\pi \times 10^{-7}~{\rm H}/{\rm m}$ is the vacuum permeability, $\mu = 1~\mu_B$ is the magnetic moment of ${}^{39}K$ atoms and $\mu_B = \frac{e \hbar}{2m_e}  = 9.274 \time 10^{-24} {\rm A~m^2}$ is the Bohr magneton.
In the following we will assume that the presence of interaction does not change the mode profile, but only generate dynamics at the level of the mode occupations.
In the case of two double-wells (that we label by $j=A, ~B$), we can simply make the four-mode approximation:
\begin{equation}
\hat{\Psi}(\mathbf{r}) =  \psi_{\rm LA}(\mathbf{r}) \hat{c}_{\rm LA} + \psi_{\rm RA}(\mathbf{r}) \hat{c}_{\rm RA} + \psi_{\rm LB}(\mathbf{r}) \hat{c}_{\rm LB} + \psi_{\rm RB}(\mathbf{r}) \hat{c}_{\rm RB}.
\end{equation}
obtained by diagonalizing separately the linear Hamiltonian for each double well.
We introduce the following spin operators 
\begin{equation}
\hat{J}^{z}_{j} = \frac{\hat{c}_{\rm Lj}^\dag \hat{c}_{\rm Lj} - \hat{c}_{\rm Rj}^\dag \hat{c}_{\rm Rj}}{2}
\end{equation}
and the total atom number in each double-well, $N_{j}$. With this, we can map each double well onto a large-S spin, with effective spin length $S_{j} = N_{j}/2$. It is possible to rewrite the interaction Hamiltonian in the following way:
\begin{align}
\label{ham_int}
\hat{H}_{\rm int} &= g I \Bigg( \left( \hat{J}^{z}_{A} \right)^{2} + \frac{\hat{N}_1^2}{4} - \frac{\hat{N}_1}{2} + \left( \hat{J}^{z}_{B} \right)^{2} + \frac{\hat{N}_2^2}{4} - \frac{\hat{N}_2}{2} \Bigg) \nonumber \\
&+ \frac{C_{dd}}{2} \Big[ 2(D - D_{\rm LR}) \hat{J}_{z{\rm 1}}^2 + 2(D - D_{\rm LR}) \hat{J}_{z{\rm 2}}^2 \nonumber \\ 
&+ \hat{N}_1  \hat{J}_{z{\rm 2}} \big( D_{\rm LALB} - D_{\rm RARB} + D_{\rm RALB} - D_{\rm LARB} \big) \nonumber \\
&+ \hat{N}_2  \hat{J}_{z{\rm 1}} \big( D_{\rm LALB} - D_{\rm RARB} - D_{\rm RALB} + D_{\rm LARB} \big) \nonumber \\
&+ 2 \hat{J}^{z}_{A}  \hat{J}^{z}_{B}  \big( D_{\rm LALB} + D_{\rm RARB} - D_{\rm RALB} - D_{\rm LARB} \big) \Big]
\end{align}
where we have introduced the dipolar integrals $D_{\alpha \beta}$, where $\alpha,\beta = ({\rm Lj,~Rk})$ with ${\rm j,~k}=A,~B$:
\begin{equation}
    D_{\alpha, \beta} \equiv \int d\mathbf{r}~  \vert \psi_{\rm \alpha}(\mathbf{r}) \vert ^{2} \int d\mathbf{r}'~   U_{dd}(\mathbf{r} - \mathbf{r}') \vert \psi_{\rm \beta}(\mathbf{r}' ) \vert ^{2}
\end{equation}
using the special notation $D \equiv D_{\alpha \alpha}$, and the contact interaction integral:
\begin{equation}
    I \equiv I_{\rm \alpha} = \int d\mathbf{r} ~ \vert \psi_{\rm \alpha}(\mathbf{r}) \vert ^{4}.
\end{equation}
Note that with these notations we assume that all the wells experience the same self interaction, which is true as long as we keep the orientation of the superlattice in the $xy$ plane. Note also that in order to obtain the Hamiltonian in Eq.~(\ref{ham_int}) we have neglected all the integrals that involve the overlap of two different wavefunctions $\psi_{\alpha} \left( \mathbf{r} \right)  \psi_{\beta} \left( \mathbf{r} \right) $ ($\alpha \neq \beta$), as the wavefunctions do not overlap in real space.
As for the linear Hamiltonian, it can be mapped onto the following operator:
\begin{equation}
\hat{H}_0 = -(E_{\rm ex} - E_{\rm gs}) \left( \hat{J}^{x}_{A} + \hat{J}^{x}_{B} \right), 
\end{equation}
where $\hat{J}^{x}_{j} = \frac{\hat{c}_{\rm Lj}^\dag \hat{c}_{\rm Rj} + \hat{c}_{\rm Lj}\hat{c}_{\rm Rj}^\dag}{2}$. Since the ground and first excited states are almost degenerate, we can neglect in the following the effect of the tunneling on the effective spin dynamics. We can rewrite the two double-well Hamiltonian in a more compact form, up to constant terms:
\begin{align}
\hat{H}_{\rm tot} &=  \hbar \Big[ \chi_{Nz}^{(AB)}  \hat{N}_A  \hat{J}^{z}_{B} + \chi_{Nz}^{(BA)}  \hat{N}_B  \hat{J}^{z}_{A} + \chi_{\rm nloc} \hat{J}^{z}_{A} \hat{J}^{z}_{B} + (\chi_{\rm loc}+\chi_{\rm cont}) \left( (\hat{J}^{z}_{A})^{2} + (\hat{J}^{z}_{B})^{2} \right) \Big]
\end{align}
where:
\begin{align}
\chi_{\rm cont} &= \frac{gI}{\hbar} \\
\chi_{\rm loc} &= \frac{C_{dd}}{\hbar} \left( D - D_{\rm LARA} \right)\\
\chi_{\rm nloc} &= \frac{C_{dd}}{\hbar } \big( D_{\rm LALB} + D_{\rm RARB} - D_{\rm RALB} - D_{\rm LARB} \big)\\
\chi_{Nz}^{(12)} &= \frac{C_{dd}}{2 \hbar } \big( D_{\rm LALB} - D_{\rm RARB} + D_{\rm RALB} - D_{\rm LARB} \big)  \\
\chi_{Nz}^{(21)} &= \frac{C_{dd}}{2 \hbar} \big( D_{\rm LALB} - D_{\rm RARB} - D_{\rm RALB} + D_{\rm LARB} \big).
\end{align}
The two local linear terms can be compensated by a spin-echo protocol, leading to the Hamiltonian given in Eq.~(1) in the main text. Moreover, the value of $\chi_{\rm loc}$ can be tuned by changing the orientation of the double well (from its dipolar contribution), while the value of the contact interaction of the atoms within a single well, $\chi_{\rm cont}$, can be tuned using a Feshbach resonance. With this, it is possible to completely cancel the local terms, and we obtain exactly the qudit Hamiltonian given in Eq.~(8) in the main text.

In the specific case of the experimental parameters presented in Ref.~\cite{PetruccianiNATCOMM2026}, with $^{39}K$ atoms, we find:
\begin{equation}
    \chi_{\rm cont} = 0.072 \frac{a}{a_{0}} {\rm Hz},  \quad \chi_{\rm loc} = -0.01 {\rm Hz}, \quad \chi_{\rm nloc} = -4.5 \times 10^{-4} {\rm Hz}\, ,
\end{equation}
where $a/a_{0}$ is the ratio between the scattering length and the Bohr radius. 
In the context of GIE experiment with $M=2$ ensembles, the time required to reach the minimum of ${\cal C}_{1}$ and ${\cal C}_{2}$ would then be $t_{min} = 1.6  \, {\rm s}$ for $N=10^{3}$ and $t_{min} = 0.4 \, {\rm s} $ for $N=4 \times 10^{3}$.
In the context of the GID experiment, with $M=2$ ensembles, the dephasing time $t_{deph}$ at which the local squeezing parameters $\xi^{2}_{\rm loc}$ goes back to $1$ would be $t_{deph} = 0.78  \,{\rm s}$ for $N=10^{3}$ and $t_{min} = 0.17  \,{\rm s}$ for $N=4 \times 10^{3}$, which is compatible with $\sim 1  \,{\rm s}$ lifetime of the BEC in this experiment.

\section*{D- Interaction Hamiltonian from mass-energy equivalence for two qudits}

Let us start by considering two qudits of length $S_{A}$ (resp. $S_{B}$), whose internal energy levels $\vert \mu \rangle$ (resp. $\vert \mu' \rangle$) are common eigenvectors of the spin operators $J^{z}_{A}$ and $\mathbf{J_{A}}^{2}$ (resp $J^{z}_{B}$ and $\mathbf{J_{2}}^{B}$), labeled by $\mu \in \left[ -S_{A}, S_{A} \right]$ (resp. $\mu' \in \left[ -S_{B}, S_{B} \right]$). 
We assume that the energy difference between two consecutive levels is given by $\Delta E$, and that it is the same for the two systems.
In the reference frame solidary to each subsystem, the local Hamiltonians of the qudits- describing free evolution in terms of their respective proper times - thus are simply given by ${\cal H}_{A} = \Delta E J^{z}_{A}$ and ${\cal H}_{B} = \Delta E J^{z}_{B}$. Starting from any product state in the form:
\begin{equation}
    \vert \psi \left( 0 \right) \rangle = \vert \psi_{A} \left( 0 \right) \rangle \otimes \vert \psi_{B} \left( 0 \right) \rangle \,, 
    \quad 
    \vert \psi_{A} \left( 0 \right) \rangle = \sum_{\mu} c^{A}_{\mu} \vert \mu \rangle \,, 
    \quad \vert \psi_{B} \left( 0 \right) \rangle = \sum_{\mu'} c^{B}_{\mu'} \vert \mu' \rangle 
    \,,
\end{equation}
and assuming the mass-energy equivalence principle, we now focus on the evolution of subsystem B in the gravitational field generated by subsystem A. 
In the solidary reference frame to subsystem B, time evolution is generated by ${\cal H}_{B}$ in terms of its proper time $\tau$ as $\frac{d}{d\tau} \vert \psi_{B} \left( \tau \right) \rangle = {\cal H}_{B}\,\vert \psi_{B} \left( \tau \right) \rangle$. From the perspective of a far-away observer, i.e. in terms of the coordinate time $t$, the state of subsystem B reads as:
\begin{equation}
    \vert \psi_{B} \left( t \right) \rangle = e^{-i\frac{t}{\hbar} \dot{\tau} {\cal H}_{B}} \vert \psi_{B} \left( 0 \right) \rangle 
    \,,
\end{equation}
where $\dot{\tau}$ is the derivative of the proper time w.r.t. the coordinate one. Following the same first order approximation as in Ref.~\cite{Castro-RuizPNAS2017}, we can write $\dot{\tau} \approx 1 + \phi(d)/c^{2}$, where $\phi(d) = - G \Delta E \mu / d c^{2}$ represents the Newtonian gravitational potential if subsystem A is in the state $\vert \mu \rangle$. The state of subsystem A thus conditions the phase acquired by subsystem B (which creates entanglement). 
If subsystem A is initially in the state $\vert \mu \rangle$, we obtain the following evolution:
\begin{align}
\vert \psi_{A} \left( t \right) \rangle &= e^{-i\frac{t}{\hbar} \Delta E \mu } \vert \mu \rangle \,\\
     \vert \psi_{B} \left( t \right) \rangle &= e^{-i\frac{t}{\hbar}  {\cal H}_{B} \left( 1 - \frac{\mu \Delta E G}{d c^{4}}\right)} \vert \psi_{B} \left( 0 \right) \rangle,
\end{align}
which, by virtue of the superposition principle, gives the global evolution:
\begin{align}
     \vert \psi \left( t \right) \rangle &= \sum_{\mu, \mu'} c^{(1)}_{\mu} c^{(2)}_{\mu'} e^{-i\frac{t}{\hbar} \Delta E \mu} e^{-i\frac{t}{\hbar} \Delta E \mu'} e^{ i\frac{t}{\hbar} \frac{\mu \mu' (\Delta E)^{2} G}{d c^{4}}}\vert \mu \rangle \otimes \vert \mu' \rangle =
      e^{-i\frac{t}{\hbar} \left( \Delta E J^{z}_{A} + \Delta E J^{z}_{B} - \frac{(\Delta E)^{2} G}{d c^{4}} J^{z}_{A}J^{z}_{B} \right)}  \vert \psi \left( 0 \right) \rangle ,
\end{align}
hence providing the same interaction Hamiltonian from Eq.~(8) of the main text, up to local term.

\section*{E- Interaction Hamiltonian from Quantum Field Theory in curved spacetime for two qudits}

As discussed in the supplementary material of Ref.~\cite{Castro-RuizPNAS2017}, a quantum clock can be viewed as a composite object emerging from the intrinsic interactions of multiple scalar fields.
Building on the framework established in Ref.~\cite{Anastopoulos2014}, this description starts from the action of two scalar fields, $\varphi_A$ and $\varphi_B$, coupled via a constant interaction matrix $M_{AB}$ in curved spacetime. In a relativistic theory, there is fundamentally no difference between mass and interaction energy. Thus, by assuming weak-gravity and taking the non-relativistic limit, the total energy of the system can be interpreted as that of a single composite system, i.e. a quantum clock field, having two internal energy levels. 
This procedure generalizes straightforwardly to $N$ matter fields, yielding a quantum clock with $N$ internal energy levels, or to a network of clocks, thereby providing the interaction Hamiltonian for a collection of gravitating quantum systems. 

Here we sketch this derivation by employing a set of $N$ interacting bosonic fields $\varphi_f$, collectively denoted by $\varphi^T =(\varphi_A,\varphi_B,...)$. The multi-component field $\varphi$ thus represents a composite quantum clock~\cite{CafassoPRD2024}, whose action is given by the sum of the free-field actions plus the interaction matrix $M$. Furthermore, given the $SO(N)$ symmetry of the massless part of the Lagrangian, we can use this freedom to pin a basis in field space where the mass matrix $M$ is diagonal from the beginning, such that $M\varphi_f = m_f\varphi_f$. Upon quantization and restriction to the two-particle sector, the effective interaction Hamiltonian~(8) presented in the Methods naturally emerges as a first-order correction in $G$. 

In curved spacetime, the coupling between the clock and the gravitational field can be effectively described starting from the sum of the Einstein–Hilbert and Klein-Gordon actions~\cite{Buchbinder2021}. In natural units $\hbar = c = 1$, this reads as
\begin{equation}
    S[g, \varphi] = \int d^4x\, \sqrt{-g}\, \left[-\frac{1}{2\kappa}R(g) +  \mathcal{L}_M(g,\varphi) \right] \,,
\end{equation}
where the gravitational coupling $\kappa:= 8\pi G $, the Lagrangian density $\mathcal{L}_M$ characterizes the clock field, and the Ricci scalar $R(g):= g^{\mu\nu}\, R_{\mu\nu}(g)$ describes the spacetime curvature in terms of the metric $g_{\mu\nu}$ - with signature $(+,-,-,-)$, and its derivatives.
In particular, the clock Lagrangian in curved spacetime reads as
\begin{equation}
    \mathcal{L}_M 
    = \frac{1}{2} \left(g^{\mu\nu} \nabla_\mu\varphi^T\nabla_\nu\varphi 
    -\varphi^TM^2\varphi\right) 
    = \frac{1}{2}\sum_f \partial_\mu \varphi_f\partial^\mu \varphi_f - \frac{1}{2}\sum_{fl}\varphi_l M^2_{lf}\varphi_f
    \,,
\end{equation}
where the covariant derivative $\nabla_\mu$ reduces to a partial derivative $\partial_\mu$ for real scalar fields, and the last term is simply $M_{lf} = m_f\delta_{lf}$.
From the stationarity condition of the action, we get
\begin{equation}
    \delta S = \int d^4x\, \sqrt{-g}\, \left[ 
    \, \frac{1}{2\kappa}\left(\kappa T_{\mu\nu} - G_{\mu\nu}\right)\,\delta g^{\mu\nu} 
    - \delta \varphi^T(\Box + M^2)\varphi
    + \nabla_\mu \Theta^\mu
    \right] = 0\,,
\end{equation}
in which the first variation provides Einstein's field equations in terms of the tensors
\begin{equation}
    G_{\mu\nu } = \frac{1}{\sqrt{-g}}\frac{\delta (\sqrt{-g}R)}{\delta g^{\mu\nu}}
    = R_{\mu\nu} - \frac{1}{2}g_{\mu\nu}R\,,
    \quad 
    T_{\mu\nu} = \frac{2}{\sqrt{-g}}
    \frac{\delta (\sqrt{-g}\mathcal{L}_M)}{\delta g^{\mu\nu}}\,
    = \partial_\mu\varphi^T\partial_\nu\varphi - g_{\mu\nu}\mathcal{L}_M 
    \,,
\end{equation}
and the second - where we introduced $\Box = \frac{1}{\sqrt{-g}}\partial_\mu \sqrt{-g}g^{\mu\nu}\partial_\nu$ - provides the equations of motion for the clock field. For our purposes, the boundary term $\nabla_\mu\Theta^\mu$ does not contribute to the physical description. Furthermore, the $\mu=\nu=0$ component of the stress-energy tensor $T_{\mu\nu}$ represents the energy density of the clock field, and is given by
\begin{equation}
    T_{00} 
    = \frac{1}{2}\partial_0\varphi^T\partial_0\varphi 
    - g_{00}g^{0j}\partial_0\varphi^T\partial_j\varphi
    - \frac{g_{00}}{2}g^{ij}\partial_i\varphi^T\partial_j\varphi 
    + \frac{g_{00}}{2} \varphi^T M^2\varphi
    \,.
\end{equation}
This object sources Einstein's field equations, giving rise to the effective description that we are seeking. 

To see that, consider the Newtonian expansion of the metric tensor, given by $g_{ij}=-\delta_{ij}$, $g_{0j} = 0$, and  $g_{00} = 1 +  2\Phi$, where the static gravitational potential $\Phi$ is first order in $\kappa$. As a consequence, we have $\sqrt{-g} = 1+\Phi$, $g^{00} \approx 1 -  2\Phi$, and $\sqrt{-g}g^{00}\approx 1-\Phi$.
Hence, the energy density of the field in the weak-gravity regime reads as
\begin{equation}
    T_0^0 = g^{00} T_{00} = 
    \frac{1}{2}\sum_f \left[\, \pi_f^2 + (\partial_i\varphi_f)^2 + (m_f \varphi_f)^2\right]
    \,,
\end{equation}
where we introduced the zero-order energy contribution $T_0^0$, and the conjugated field $\pi_f = \sqrt{-g}g^{00}\partial_0\varphi_f$. Moreover, we introduce the first-order clock Hamiltonian as
\begin{equation}
    H = \int d^3 x \sqrt{-g}T_{0}^{\,0} = \frac{1}{2}\int d^3 x\,(1 + \Phi)\sum_f \left[\, \pi_f^2 + (\partial_i\varphi_f)^2 + (m_f \varphi_f)^2\right] \,,
\end{equation}
in which, evaluating the gravitational potential $\Phi$ requires solving Einstein's field equations as
\begin{equation}
    G_{0}^{\,0} - \kappa T_{0}^{\,0} = -2\nabla^2\Phi - \kappa\,T_{0}^{\,0}=0  
    \implies 
    \Phi(x) = -\frac{\kappa}{8 \pi} \int d^3x' \,\frac{T_{0}^{\,0}(x')}{|x-x'|} \,.
\end{equation}

To conclude this derivation, we now proceed to quantize the clock field and perform the non-relativistic limit. The equations of motion in the Newtonian regime read as
\begin{equation}
    0=[\Box + M^2]\varphi  
    = \left[g^{\mu\nu}\partial_\mu\partial_\nu +\frac{\partial_\mu(\sqrt{-g}\,g^{\mu\nu})}{\sqrt{-g}}\partial_\nu +M^2 \right] \varphi 
    \approx \left[(1-2\Phi)\partial_0^2 - \partial_i^2 + M^2 \right] \varphi
    \,,
\end{equation}
where we neglected the terms proportional to $\partial_j \Phi$. Rewriting them as
\begin{equation}
     \partial_0^2\varphi - \partial_i^2\varphi + M^2\varphi = -2\Phi(M^2- \partial_i^2)\varphi \,,
\end{equation}
we notice that the zero-order solution suffices to compute the first-order correction to the Hamiltonian. Then, following Ref.~\cite{Castro-RuizPNAS2017}, the quantized solutions in the non-relativistic limit, and at zero-order in $\kappa$, read as
\begin{equation}
    \varphi_f \approx \frac{1}{\sqrt{2m_f}}\left(\chi_f + \chi^\dag_f\right)\,,
    \quad 
    \pi_f \approx i\sqrt{\frac{m_f}{2}}\left(\chi_f - \chi^\dag_f\right)\,,
    \quad \text{with }\,
    \chi_f = \int \frac{d^3k}{(2\pi)^3}e^{i(m_ft - kx)}{a}_{fk}\,,
\end{equation}
where the operator ${a}_{fk}$ satisfies $\left[\,{a}_{fk}\,,\,{a}_{f'k'}^\dag\right] = \delta_{ff'}\delta^3(k-k')$. In terms of $\chi$, we thus obtain
\begin{equation}
    T_{0}^{\,0} = \chi^\dag M \chi - \frac{1}{2}\chi^\dag M^{-1}\nabla^2\chi \,,
\end{equation}
and the Hamiltonian for slow moving particles is then given by
\begin{equation}
    H =\int d^3 x\, \chi^\dag  M \chi - \int d^3x\,d^3x'\,\chi^\dag(x)\,\chi^\dag(x')  \frac{GM^2}{|x-x'|} \,\chi(x)\,\chi(x')\,,
\end{equation}
from which the interaction Hamiltonian employed in the main text arises. In particular, one can identify $M \equiv 1m + \hat{J}^z\Delta E$, such that $m_f = m + j_f\Delta E  $, and following the same steps of Ref.s~\cite{Anastopoulos2014,Castro-RuizPNAS2017} to restrict to the two-particle sector, the interaction term can be derived from the matrix elements of H as
\begin{equation}
    H_{int} = - \frac{G(\Delta E)^2}{|x_A - x_B|} \hat{J}^z_A \otimes \hat{J}^z_B\,,
\end{equation}
which has the same form as Eq.~(8) of the Methods. Thus, describing quantum clocks as non-relativistic particles,
the mass-energy equivalence prescribes this gravitational-like interaction to take place. Finally, the effect of self-interaction in the one-particle sector gives rise to a mass-renormalization term~\cite{Castro-RuizPNAS2017}, whose quantum contribution to the entanglement goes beyond the scope of this paper.

\section*{F- Analytic formulae for $M=3$ and $4$}

In order to compute the time evolution of the squeezing and the covariance matrix, we use the following set of analytical formulae, computed for an arbitrary time evolution of duration $t$ under the general Hamiltonian given by:
\begin{equation}
    H_{\rm net} = (\chi_{\rm cont} + \chi_{\rm loc})\sum_{i=1}^{M} (J^{z}_{i})^{2} + \chi_{\rm nloc} \sum_{i < j} J^{z}_{i} J^{z}_{j}. 
\end{equation}
We assume we have $M$ subsystems each made of a $N$-particle bimodal BEC (equivalent to an effective $N+1$-level qudit), and the initial state is a product state of coherent spin states aligned along $x$, $\vert \psi (t=0) \rangle = \bigotimes_{i = 1}^{M} \vert CSS_{x} \rangle_{i}$. First, we give the values for the diagonal blocks, which involve only local correlations:
\begin{equation}
    \langle J^{x}_{i} \rangle (t) = \frac{N}{2}  \cos \left(  (\chi_{\rm cont}+\chi_{\rm loc}) t \right) ^{N-1} \cos \left( \frac{\chi_{\rm nloc}}{2}  t \right) ^{N(M-1)}
\end{equation}
\begin{equation}
    \langle J^{y}_i \rangle (t) = 0, \quad  \langle J^{z}_i \rangle (t) = 0
\end{equation}
\begin{equation}
    \langle (J^{x}_{i})^{2} \rangle (t) = \frac{N\left( N + 1 \right)}{8} +   \frac{N\left( N - 1 \right)}{8}  \cos \left( 2  (\chi_{\rm cont}+\chi_{\rm loc})  t \right)^{N-2}  \cos \left( \chi_{\rm nloc} t \right)^{N(M-1)} 
\end{equation}
\begin{equation}
    \langle (J^{y}_{i})^{2} \rangle (t) =  \frac{N\left( N + 1 \right)}{8}  -  \frac{N\left( N - 1 \right)}{8}  \cos \left( 2  (\chi_{\rm cont}+\chi_{\rm loc})  t \right)^{N-2}  \cos \left( \chi_{\rm nloc} t \right)^{N(M-1)} 
\end{equation}
\begin{equation}
    \langle (J^{z}_{i})^{2} \rangle (t) = N/4
\end{equation}

\begin{equation}
    \langle J^{x}_{i} J^{y}_{i} \rangle (t) =  \langle J^{y}_{i} J^{x}_{i} \rangle (t) =  \langle J^{x}_{i} J^{z}_{i} \rangle (t) = \langle J^{z}_{i} J^{x}_{i} \rangle (t) = 0
\end{equation}
\begin{equation}
    \langle J^{y}_{i} J^{z}_{i} \rangle (t) = \langle J^{y}_{i} J^{z}_{i} \rangle (t) =  \frac{N \left(N - 1 \right)}{4} \cos \left( (\chi_{\rm cont}+\chi_{\rm loc})  t \right)^{N-2} \sin \left(  (\chi_{\rm cont}+\chi_{\rm loc}) t \right)  \cos \left( \frac{\chi_{\rm nloc}}{2}  t \right) ^{N (M-1)} 
\end{equation}
Finally, we give the formulae for the off block-diagonal elements of the covariance matrix $\Gamma$:
\begin{equation}
    \langle J^{x}_{i} J^{x}_{j} \rangle (t) =  \frac{N^2}{4} \cos \left(  \left( \chi_{\rm cont} + \chi{\rm loc} - \frac{\chi_{\rm nloc}}{2} \right) t \right)^{2N -2} + \frac{N^2}{4} \cos \left(  \left( \chi_{\rm cont} + \chi{\rm loc} + \frac{\chi_{\rm nloc}}{2} \right) t \right)^{2N -2} \cos \left( \chi_{\rm nloc} t \right)^{N(M-2)}
\end{equation}
\begin{equation}
    \langle J^{y}_{i} J^{y}_{j} \rangle (t) =  \frac{N^2}{4} \cos \left(  \left( \chi_{\rm cont} + \chi{\rm loc} - \frac{\chi_{\rm nloc}}{2} \right) t \right)^{2N -2} - \frac{N^2}{4} \cos \left(  \left( \chi_{\rm cont} + \chi{\rm loc} + \frac{\chi_{\rm nloc}}{2} \right) t \right)^{2N -2} \cos \left( \chi_{\rm nloc} t \right)^{N(M-2)}
\end{equation}
\begin{equation}
    \langle J^{z}_{i} J^{z}_{j} \rangle (t) =  0
\end{equation}

\begin{equation}
    \langle J^{x}_{i} J^{y}_{j} \rangle (t) =  \langle J^{x}_{i} J^{z}_{j} \rangle (t) =  0
\end{equation}
\begin{equation}
    \langle J^{y}_{i} J^{z}_{j} \rangle (t)  =  \frac{N}{4} \sin \left( \frac{\chi_{\rm nloc}}{2} t \right) \cos \left(  \frac{\chi_{\rm nloc}}{2} t \right)^{N-1}   \cos \left( (\chi_{\rm cont} + \chi_{\rm loc} ) t \right)^{N-1} \cos \left( \frac{\chi_{\rm nloc}}{2}  t \right) ^{N(M-2)}
\end{equation}

\section*{G- Semi-analytic approach for the GID experiment}

The experimental procedure described in Fig.~3 consists in evolving an initial product of coherent spin state along $x$ first with a local OAT Hamiltonian $H_{OAT} = (\chi_{\rm cont} + \chi_{\rm loc} )\sum_{i} (J^{z}_{i})^{2}$ for a time $t_{\rm rot}$, then locally rotating the state by an angle $\theta$ around $x$ axis, and finally evolve with the interaction Hamiltonian $\chi_{\rm nloc} J^{z}_{A} \otimes J^{z}_{B}$ for a time $t-t_{\rm rot}$.

In the Heisenberg picture, the resulting spin observables are given by:
\begin{align}
    J^{+}_{A} (t_{\rm rot}, \theta, t-t_{\rm rot}) &= \exp [ i \chi_{\rm nloc} (t-t_{\rm rot}) ( \cos(\theta) J^{z}_{B} + \sin ( \theta ) J^{y}_{B} (t_{\rm rot}) )] (J^{x}_{A} (t_{\rm rot}) + i \cos \theta J^{y}_{A} (t_{\rm rot}) - \sin \theta J^{z}_{A}) \\
    J^{z}_{A} (t_{\rm rot}, \theta, t-t_{\rm rot}) &= \cos(\theta) J^{z}_{A} + \sin ( \theta ) J^{y}_{A} (t_{\rm rot}) \\
    J^{+}_{B} (t_{\rm rot}, \theta, t-t_{\rm rot}) &= \exp [ i \chi_{\rm nloc} (t-t_{\rm rot}) ( \cos(\theta) J^{z}_{A} + \sin ( \theta ) J^{y}_{A} (t_{\rm rot}) )] (J^{x}_{B} (t_{\rm rot}) + i \cos \theta J^{y}_{B} (t_{\rm rot}) - \sin \theta J^{z}_{B}) \\
    J^{z}_{B} (t_{\rm rot}, \theta, t-t_{\rm rot}) &= \cos(\theta) J^{z}_{B} + \sin ( \theta ) J^{y}_{B} (t_{\rm rot}) 
\end{align}
with
\begin{align}
    J^{x}_{j} (t_{\rm rot}) &= (J^{+}_{j}(t_{\rm rot}) + J^{-}_{j}(t_{\rm rot}))/2\\
    J^{y}_{j} (t_{\rm rot}) &= (J^{+}_{j}(t_{\rm rot}) - J^{-}_{j}(t_{\rm rot}))/2i\\
    J^{+}_{j} (t_{\rm rot}) &= J^{+}_{j} \exp[i(\chi_{\rm cont}+\chi_{\rm loc})t_{\rm rot}(2J^{z}_{j} + 1)] \\
    J^{-}_{j} (t_{\rm rot}) &=  \exp[-i(\chi_{\rm cont}+\chi_{\rm loc})t_{\rm rot}(2J^{z}_{j} + 1)] J^{-}_{j}
\end{align}
and $j = A, B$. In the case where $\theta = 0$, we can compute analytically any mean value and two point correlation functions of these observables. However, for $\theta \neq 0$, computing the mean value of $\langle CSS_{x} \vert \exp [ i \chi_{\rm nloc} (t-t_{\rm rot}) ( \cos(\theta) J^{z}_{j} + \sin ( \theta ) J^{y}_{j} (t_{\rm rot}) )] \vert CSS_{x} \rangle_{j} $ on subsystem $j$ becomes impossible. Yet, since our initial state is a product state, and since we can separate the contribution of subsystems $A$ and $B$ in the Heisenberg evolution of the observables, we can numerically compute any correlation function using only local mean values (requiring vectors of length N+1 and matrices of size $(N+1)\times(N+1)$, and we never need to store and use the wavevector of the total system (of length $(N+1)^2$) nor observables (of size $(N+1)^2\times(N+1)^2$). 
For this purpose, we first introduce the following local quantity:
\begin{equation}
    V_{z} (t_{\rm rot}, \theta, t-t_{\rm rot}) = \langle \psi (t_{\rm rot}) \vert  \exp [ i \chi_{\rm nloc} (t-t_{\rm rot})\tilde{J}^{z} (\theta)] \vert \psi (t_{\rm rot}) \rangle 
\end{equation}
with the notation $\tilde{J}^{z} (\theta) =  \cos(\theta) J^{z} + \sin ( \theta ) J^{y}$. Note that we take the mean value on the local state evolved with a local OAT dynamics for duration $t_{\rm rot}$, and we removed the subscript $j$ as at this stage all the subsystems are equivalent. In a similar fashion, we introduce
\begin{align}
    V_{az} (t_{\rm rot}, \theta, t-t_{\rm rot}) &= \langle \psi (t_{\rm rot}) \vert \tilde{J}^{a} (\theta)  \exp [ i \chi_{\rm nloc} (t-t_{\rm rot})\tilde{J}^{z} (\theta)] \vert \psi (t_{\rm rot}) \rangle \\
     V_{za} (t_{\rm rot}, \theta, t-t_{\rm rot}) &= \langle \psi (t_{\rm rot}) \vert  \exp [ i \chi_{\rm nloc} (t-t_{\rm rot})\tilde{J}^{z} (\theta)] \tilde{J}^{a} (\theta) \vert \psi (t_{\rm rot}) \rangle
\end{align}
with $a = x, y, z$ and $\tilde{J}^{x} (\theta) = J^{x}$, $\tilde{J}^{y} (\theta) = \cos(\theta) J^{y} - \sin ( \theta ) J^{z}$ and their respective real and imaginary part that we call $R_{az}, R_{za}$ and $I_{az}, I_{za}$ respectively. In the case $a=z$, the notation may seem redundant but the two quantities are indeed equal.
From the different commutation relations between the spin operators, we can derive the following properties:
\begin{align}
    R_{xz}(t_{\rm rot}, \theta, t- t_{\rm rot}) &= R_{xz} (t_{\rm rot}, \theta, -(t- t_{\rm rot})) = R_{zx} (t_{\rm rot}, \theta, t- t_{\rm rot}) = R_{zx} (t_{\rm rot}, \theta, -(t- t_{\rm rot})) \\ 
    I_{xz}(t_{\rm rot}, \theta, t- t_{\rm rot}) &= I_{xz} (t_{\rm rot}, \theta, -(t- t_{\rm rot})) = -I_{zx} (t_{\rm rot}, \theta, t- t_{\rm rot}) = -I_{zx} (t_{\rm rot}, \theta, -(t- t_{\rm rot})) \\ 
    R_{yz}(t_{\rm rot}, \theta, t- t_{\rm rot}) &= -R_{yz} (t_{\rm rot}, \theta, -(t- t_{\rm rot})) = -R_{zy} (t_{\rm rot}, \theta, t- t_{\rm rot}) = R_{zy} (t_{\rm rot}, \theta, -(t- t_{\rm rot})) \\ 
    I_{yz}(t_{\rm rot}, \theta, t- t_{\rm rot}) &= -I_{yz} (t_{\rm rot}, \theta, -(t- t_{\rm rot})) = I_{zy} (t_{\rm rot}, \theta, t- t_{\rm rot}) = -I_{zy} (t_{\rm rot}, \theta, -(t- t_{\rm rot}))\\
    R_{zz}(t_{\rm rot}, \theta, t- t_{\rm rot}) &= R_{zz} (t_{\rm rot}, \theta, -(t- t_{\rm rot})) = 0\\
    I_{zz} (t_{\rm rot}, \theta, t- t_{\rm rot}) &= - I_{zz} (t_{\rm rot}, \theta, -(t- t_{\rm rot}))
\end{align}
With this, we can rewrite any mean value and two point correlation function (with again $\tau_{\rm rot} = (\chi_{\rm cont}+\chi_{\rm loc}) t_{\rm rot}$:
\begin{equation}
    \langle J^{x}_{i} \rangle (t_{\rm rot}, \theta, t- t_{\rm rot}) = \frac{N}{2}  \cos \left( \tau_{\rm rot} \right) ^{N-1} V_{z} (t_{\rm rot}, \theta, t- t_{\rm rot})
\end{equation}
\begin{equation}
    \langle J^{y}_i \rangle (t_{\rm rot}, \theta, t- t_{\rm rot}) = 0, \quad  \langle J^{z}_i \rangle (t_{\rm rot}, \theta, t- t_{\rm rot})= 0
\end{equation}
\begin{align}
    \langle (J^{x}_{i})^{2} \rangle (t_{\rm rot}, \theta, t- t_{\rm rot}) &= \left[ \frac{N\left( N + 1 \right)}{16} +   \frac{N\left( N - 1 \right)}{16}  \cos \left( 2  \tau_{rot} \right)^{N-2}\right] (1+V_{z}(t_{\rm rot}, \theta, 2(t- t_{\rm rot})) \nonumber\\
    &+\cos (\theta)^{2}  \left[ \frac{N\left( N + 1 \right)}{16}  -  \frac{N\left( N - 1 \right)}{16}  \cos \left( 2  \tau_{\rm rot} \right)^{N-2}\right] (1-V_{z}(t_{\rm rot}, \theta, 2(t- t_{\rm rot}))\nonumber\\
    &+ \sin (\theta)^{2} \frac{N}{8} (1(V_{z}(t_{\rm rot}, \theta, 2(t- t_{\rm rot}))\nonumber\\
    &+ \cos (\theta) \sin (\theta) \frac{N(N-1)}{4} \cos \left(  \tau_{\rm rot} \right)^{N-2} \sin\left(  \tau_{\rm rot}\right) (V_{z}(t_{\rm rot}, \theta, 2(t- t_{\rm rot})-1)
\end{align}
\begin{align}
    \langle (J^{y}_{i})^{2} \rangle (t_{\rm rot}, \theta, t- t_{\rm rot}) &= \left[ \frac{N\left( N + 1 \right)}{16} +   \frac{N\left( N - 1 \right)}{16}  \cos \left( 2  \tau_{\rm rot} \right)^{N-2}\right] (1-V_{z}(t_{\rm rot}, \theta, 2(t- t_{\rm rot}))\nonumber \\
    &+\cos (\theta)^{2}  \left[ \frac{N\left( N + 1 \right)}{16}  -  \frac{N\left( N - 1 \right)}{16}  \cos \left( 2  \tau_{\rm rot} \right)^{N-2}\right] (1+V_{z}(t_{\rm rot}, \theta, 2(t- t_{\rm rot}))\nonumber\\
    &+ \sin (\theta)^{2} \frac{N}{8} (1 + V_{z}(t_{\rm rot}, \theta, 2(t- t_{\rm rot}))\nonumber\\
    &- \cos (\theta) \sin (\theta) \frac{N(N-1)}{4} \cos \left( \tau_{\rm rot} \right)^{N-2} \sin\left(  \tau_{\rm rot} \right) (1+V_{z}(t_{\rm rot}, \theta, 2(t- t_{\rm rot}))
\end{align}
\begin{align}
    \langle (J^{z}_{i})^{2} \rangle (t_{\rm rot}, \theta, t- t_{\rm rot}) &= \cos (\theta)^{2} \frac{N}{4} + \sin (\theta)^{2} \left[\frac{N\left( N + 1 \right)}{8}  -  \frac{N\left( N - 1 \right)}{8}  \cos \left( 2 \tau_{\rm rot} \right)^{N-2} \right]\nonumber\\
    &+ \cos (\theta) \sin (\theta)  \frac{N(N-1)}{2} \cos \left( \tau_{\rm rot} \right)^{N-2} \sin\left( \tau_{\rm rot}\right)
\end{align}
\begin{equation}
    \langle J^{x}_{i} J^{y}_{i} \rangle (t_{\rm rot}, \theta, t- t_{\rm rot}) =  \langle J^{y}_{i} J^{x}_{i} \rangle (t_{\rm rot}, \theta, t- t_{\rm rot}) =  \langle J^{x}_{i} J^{z}_{i} \rangle (t_{\rm rot}, \theta, t- t_{\rm rot}) = \langle J^{z}_{i} J^{x}_{i} \rangle (t_{\rm rot}, \theta, t- t_{\rm rot}) = 0
\end{equation}
\begin{align}
    \langle \lbrace J^{y}_{i}, ~J^{z}_{i} \rbrace \rangle (t_{\rm rot}, \theta, t- t_{\rm rot}) &= V_{z} (t_{\rm rot}, \theta, t- t_{\rm rot}) \cos\left(2\theta \right) \frac{N \left(N - 1 \right)}{2} \cos \left( \tau_{\rm rot}\right)^{N-2} \sin \left( \tau_{\rm rot} \right) \nonumber\\
    &+ V_{z} (t_{\rm rot}, \theta, t- t_{\rm rot})  \sin ( 2 \theta)  \left[  \frac{N\left( N + 1 \right)}{8}  -  \frac{N\left( N - 1 \right)}{8}  \cos \left( 2  \tau_{\rm rot}\right)^{N-2} - \frac{N}{4} \right]  
\end{align}
\begin{align}
    \langle J^{x}_{A} J^{x}_{B} \rangle (t_{\rm rot}, \theta, t- t_{\rm rot}) &= \frac{\left[R_{xz}(t_{\rm rot}, \theta, t- t_{\rm rot}) + I_{yz}(t_{\rm rot}, \theta, t- t_{\rm rot}) \right]^{2} + \left[I_{xz}(t_{\rm rot}, \theta, t- t_{\rm rot}) - R_{yz}(t_{\rm rot}, \theta, t- t_{\rm rot}) \right]^{2}}{2} \nonumber \\
    &+ \frac{\left[R_{xz}(t_{\rm rot}, \theta, t- t_{\rm rot}) - I_{yz}(t_{\rm rot}, \theta, t- t_{\rm rot}) \right]^{2} + \left[I_{xz}(t_{\rm rot}, \theta, t- t_{\rm rot}) - R_{yz}(t_{\rm rot}, \theta, t- t_{\rm rot}) \right]^{2}}{2} 
\end{align}
\begin{equation}
    \langle J^{y}_{A} J^{y}_{B} \rangle  (t_{\rm rot}, \theta, t- t_{\rm rot}) =  2\left[ R_{xz}  (t_{\rm rot}, \theta, t- t_{\rm rot}) I_{yz}  (t_{\rm rot}, \theta, t- t_{\rm rot}) -  R_{yz}  (t_{\rm rot}, \theta, t- t_{\rm rot}) I_{xz} (t_{\rm rot}, \theta, t- t_{\rm rot})\right]
\end{equation}
\begin{equation}
    \langle J^{z}_{i} J^{z}_{j} \rangle  (t_{\rm rot}, \theta, t- t_{\rm rot})=  0
\end{equation}
\begin{equation}
    \langle J^{x}_{i} J^{y}_{j} \rangle  (t_{\rm rot}, \theta, t- t_{\rm rot})=  \langle J^{x}_{i} J^{z}_{j} \rangle  (t_{\rm rot}, \theta, t- t_{\rm rot}) =  0
\end{equation}
\begin{equation}
    \langle \lbrace J^{y}_{A}, ~J^{z}_{B} \rbrace \rangle  (t_{\rm rot}, \theta, t- t_{\rm rot}) = \langle \lbrace J^{y}_{B}, ~J^{z}_{A} \rbrace \rangle  (t_{\rm rot}, \theta, t- t_{\rm rot}) = N \cos \left( \tau_{\rm rot} \right)^{N-1} I_{zz} (t_{\rm rot}, \theta, t- t_{\rm rot})
\end{equation}

where we evaluated numerically the values of the $V_{z}, R_{az}, R_{za}, I_{az} I_{za}$ and all the rest is obtained analytically. This trick allows us to significantly increase the size of the system we can study, from $N=100$ to $N>1000$. Moreover, this can be easily generalized to the case of an arbitrary number $M$ of subsystems, while requiring the same computational time as $M=2$, while we would expect an exponential scaling with $M$ for the naive approach:
\begin{equation}
    \langle J^{x}_{i} \rangle (t_{\rm rot}, \theta, t- t_{\rm rot}) = \frac{N}{2}  \cos \left( \tau_{\rm rot} \right) ^{N-1} V_{z} (t_{\rm rot}, \theta, t- t_{\rm rot})^{M-1}
\end{equation}
\begin{equation}
    \langle J^{y}_i \rangle (t_{\rm rot}, \theta, t- t_{\rm rot}) = 0, \quad  \langle J^{z}_i \rangle (t_{\rm rot}, \theta, t- t_{\rm rot})= 0
\end{equation}
\begin{align}
    \langle (J^{x}_{i})^{2} \rangle (t_{\rm rot}, \theta, t- t_{\rm rot}) &= \left[ \frac{N\left( N + 1 \right)}{16} +   \frac{N\left( N - 1 \right)}{16}  \cos \left( 2  \tau_{rot} \right)^{N-2}\right] (1+V_{z}(t_{\rm rot}, \theta, 2(t- t_{\rm rot}))^{M-1}) \nonumber\\
    &+\cos (\theta)^{2}  \left[ \frac{N\left( N + 1 \right)}{16}  -  \frac{N\left( N - 1 \right)}{16}  \cos \left( 2  \tau_{\rm rot} \right)^{N-2}\right] (1-V_{z}(t_{\rm rot}, \theta, 2(t- t_{\rm rot}))^{M-1})\nonumber\\
    &+ \sin (\theta)^{2} \frac{N}{8} (1-V_{z}(t_{\rm rot}, \theta, 2(t- t_{\rm rot}))^{M-1})\nonumber\\
    &+ \cos (\theta) \sin (\theta) \frac{N(N-1)}{4} \cos \left(  \tau_{\rm rot} \right)^{N-2} \sin\left(  \tau_{\rm rot}\right) (V_{z}(t_{\rm rot}, \theta, 2(t- t_{\rm rot}))^{M-1}-1)
\end{align}
\begin{align}
    \langle (J^{y}_{i})^{2} \rangle (t_{\rm rot}, \theta, t- t_{\rm rot}) &= \left[ \frac{N\left( N + 1 \right)}{16} +   \frac{N\left( N - 1 \right)}{16}  \cos \left( 2  \tau_{\rm rot} \right)^{N-2}\right] (1-V_{z}(t_{\rm rot}, \theta, 2(t- t_{\rm rot}))^{M-1})\nonumber \\
    &+\cos (\theta)^{2}  \left[ \frac{N\left( N + 1 \right)}{16}  -  \frac{N\left( N - 1 \right)}{16}  \cos \left( 2  \tau_{\rm rot} \right)^{N-2}\right] (1+V_{z}(t_{\rm rot}, \theta, 2(t- t_{\rm rot}))^{M-1})\nonumber\\
    &+ \sin (\theta)^{2} \frac{N}{8} (1 + V_{z}(t_{\rm rot}, \theta, 2(t- t_{\rm rot}))^{M-1})\nonumber\\
    &- \cos (\theta) \sin (\theta) \frac{N(N-1)}{4} \cos \left( \tau_{\rm rot} \right)^{N-2} \sin\left(  \tau_{\rm rot} \right) (1+V_{z}(t_{\rm rot}, \theta, 2(t- t_{\rm rot}))^{M-1})
\end{align}
\begin{align}
    \langle (J^{z}_{i})^{2} \rangle (t_{\rm rot}, \theta, t- t_{\rm rot}) &= \cos (\theta)^{2} \frac{N}{4} + \sin (\theta)^{2} \left[\frac{N\left( N + 1 \right)}{8}  -  \frac{N\left( N - 1 \right)}{8}  \cos \left( 2 \tau_{\rm rot} \right)^{N-2} \right]\nonumber\\
    &+ \cos (\theta) \sin (\theta)  \frac{N(N-1)}{2} \cos \left( \tau_{\rm rot} \right)^{N-2} \sin\left( \tau_{\rm rot}\right)
\end{align}
\begin{equation}
    \langle J^{x}_{i} J^{y}_{i} \rangle (t_{\rm rot}, \theta, t- t_{\rm rot}) =  \langle J^{y}_{i} J^{x}_{i} \rangle (t_{\rm rot}, \theta, t- t_{\rm rot}) =  \langle J^{x}_{i} J^{z}_{i} \rangle (t_{\rm rot}, \theta, t- t_{\rm rot}) = \langle J^{z}_{i} J^{x}_{i} \rangle (t_{\rm rot}, \theta, t- t_{\rm rot}) = 0
\end{equation}
\begin{align}
    \langle \lbrace J^{y}_{i}, ~J^{z}_{i} \rbrace \rangle (t_{\rm rot}, \theta, t- t_{\rm rot}) &= V_{z} (t_{\rm rot}, \theta, t- t_{\rm rot})^{M-1} \cos\left(2\theta \right) \frac{N \left(N - 1 \right)}{2} \cos \left( \tau_{\rm rot}\right)^{N-2} \sin \left( \tau_{\rm rot} \right) \nonumber\\
    &+ V_{z} (t_{\rm rot}, \theta, t- t_{\rm rot})^{M-1}  \sin ( 2 \theta)  \left[  \frac{N\left( N + 1 \right)}{8}  -  \frac{N\left( N - 1 \right)}{8}  \cos \left( 2  \tau_{\rm rot}\right)^{N-2} - \frac{N}{4} \right]  
\end{align}
\begin{align}
    \langle J^{x}_{i} J^{x}_{j} \rangle (t_{\rm rot}, \theta, t- t_{\rm rot}) &= \frac{\left[R_{xz}(t_{\rm rot}, \theta, t- t_{\rm rot}) + I_{yz}(t_{\rm rot}, \theta, t- t_{\rm rot}) \right]^{2} + \left[I_{xz}(t_{\rm rot}, \theta, t- t_{\rm rot}) - R_{yz}(t_{\rm rot}, \theta, t- t_{\rm rot}) \right]^{2}}{2} \nonumber \\
    &+ V_{z} (t_{\rm rot}, \theta, 2(t- t_{\rm rot})) ^{M-2} \frac{\left[R_{xz}(t_{\rm rot}, \theta, t- t_{\rm rot}) - I_{yz}(t_{\rm rot}, \theta, t- t_{\rm rot}) \right]^{2} + \left[I_{xz}(t_{\rm rot}, \theta, t- t_{\rm rot}) - R_{yz}(t_{\rm rot}, \theta, t- t_{\rm rot}) \right]^{2}}{2} 
\end{align}
\begin{align}
    \langle J^{y}_{i} J^{y}_{j} \rangle  (t_{\rm rot}, \theta, t- t_{\rm rot}) &=  \frac{\left[R_{xz}(t_{\rm rot}, \theta, t- t_{\rm rot}) + I_{yz}(t_{\rm rot}, \theta, t- t_{\rm rot}) \right]^{2} + \left[I_{xz}(t_{\rm rot}, \theta, t- t_{\rm rot}) - R_{yz}(t_{\rm rot}, \theta, t- t_{\rm rot}) \right]^{2}}{2} \nonumber \\
    &- V_{z} (t_{\rm rot}, \theta, 2(t- t_{\rm rot})) ^{M-2} \frac{\left[R_{xz}(t_{\rm rot}, \theta, t- t_{\rm rot}) - I_{yz}(t_{\rm rot}, \theta, t- t_{\rm rot}) \right]^{2} + \left[I_{xz}(t_{\rm rot}, \theta, t- t_{\rm rot}) - R_{yz}(t_{\rm rot}, \theta, t- t_{\rm rot}) \right]^{2}}{2} 
\end{align}
\begin{equation}
    \langle J^{z}_{i} J^{z}_{j} \rangle  (t_{\rm rot}, \theta, t- t_{\rm rot})=  0
\end{equation}
\begin{equation}
    \langle J^{x}_{i} J^{y}_{j} \rangle  (t_{\rm rot}, \theta, t- t_{\rm rot})=  \langle J^{x}_{i} J^{z}_{j} \rangle  (t_{\rm rot}, \theta, t- t_{\rm rot}) =  0
\end{equation}
\begin{equation}
    \langle \lbrace J^{y}_{i}, ~J^{z}_{j} \rbrace \rangle  (t_{\rm rot}, \theta, t- t_{\rm rot}) = N \cos \left( \tau_{\rm rot} \right)^{N-1} I_{zz} (t_{\rm rot}, \theta, t- t_{\rm rot}) V_{z} (t_{\rm rot}, \theta, t- t_{\rm rot}) ^{M-2}
\end{equation}

\section*{H- Generalization of the GID experiment for M=3 and M=4}

\begin{figure*}[h!]
\includegraphics[width=0.45\textwidth]{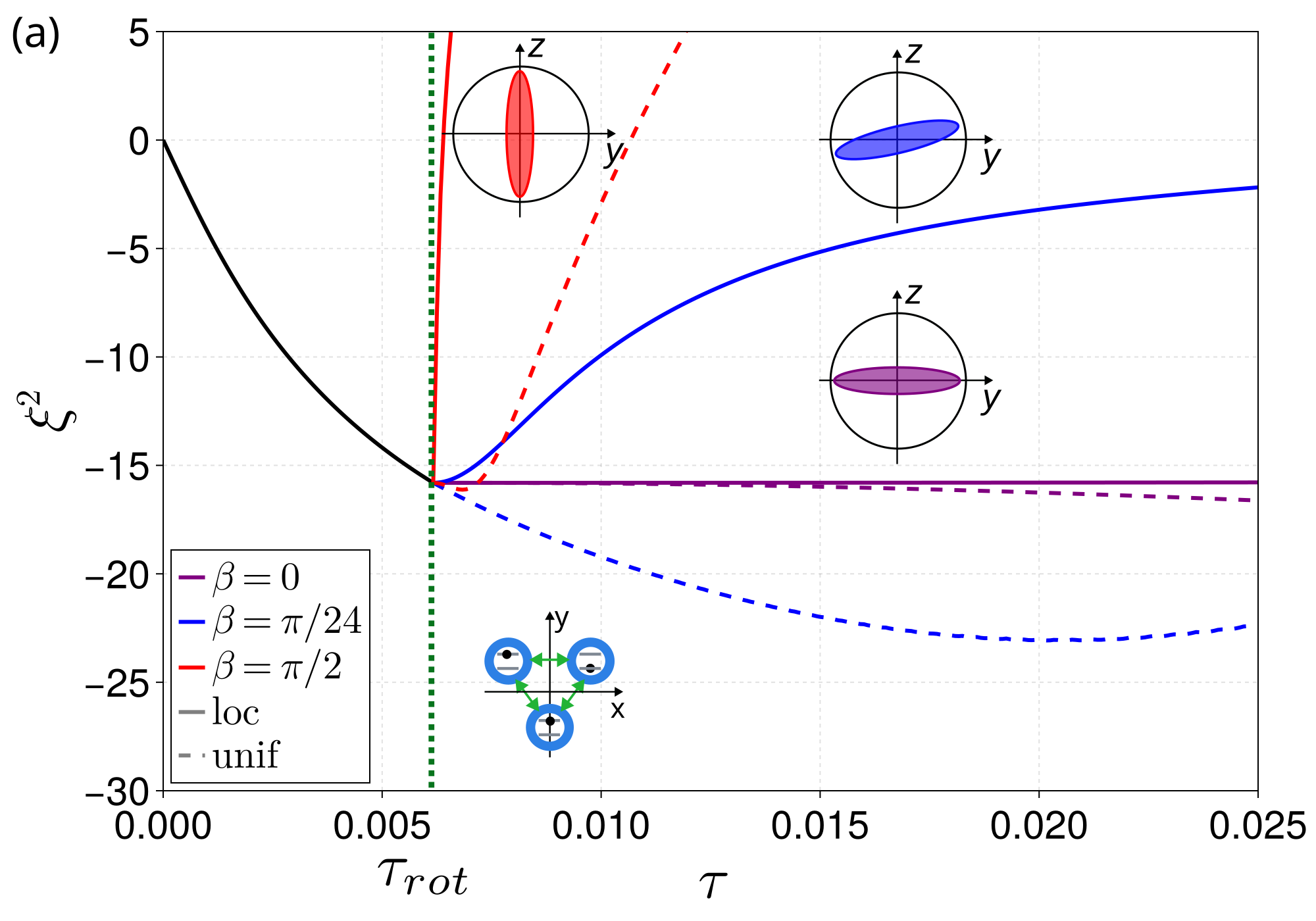}
\includegraphics[width=0.45\textwidth]{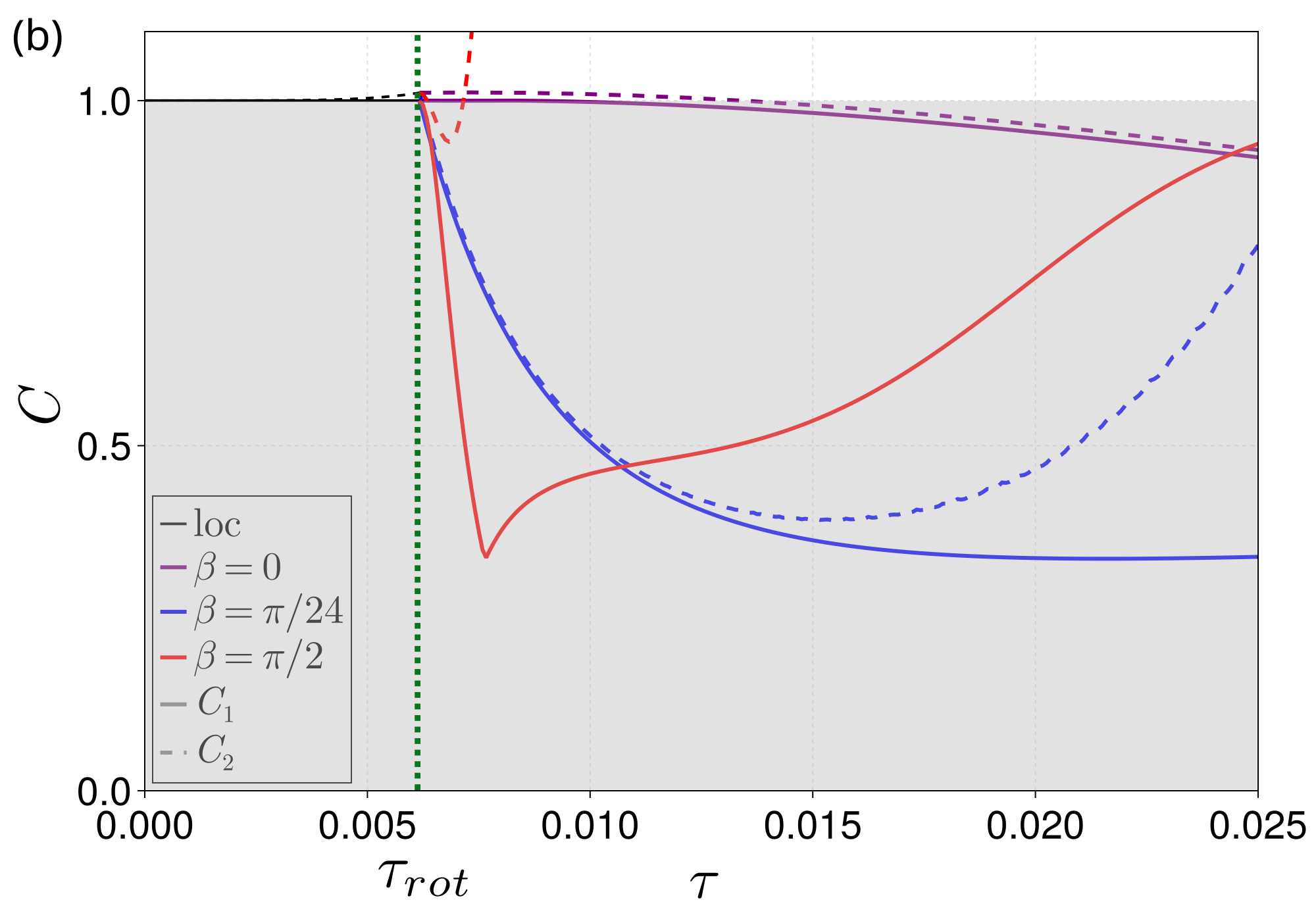}
\includegraphics[width=0.45\textwidth]{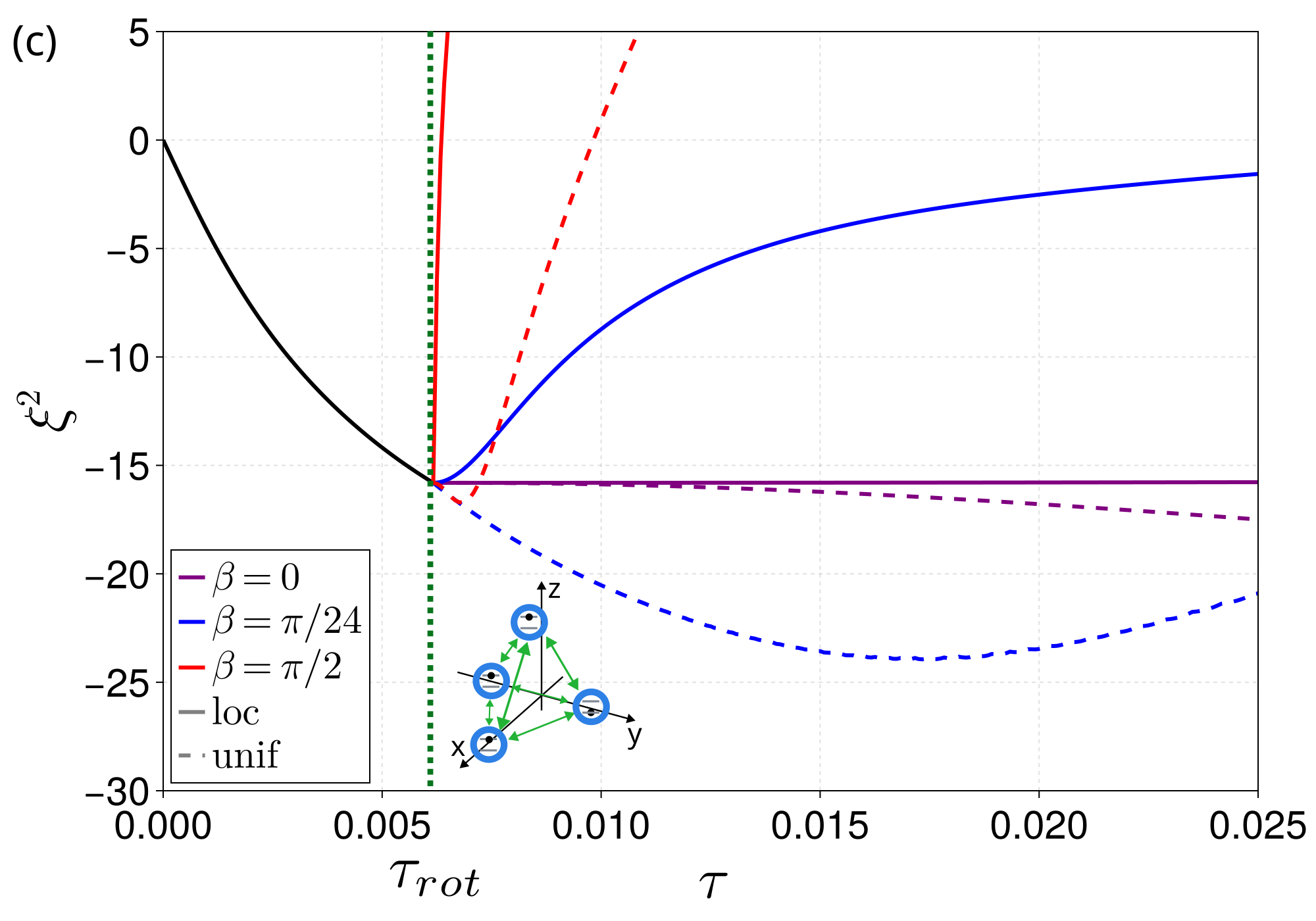}
\includegraphics[width=0.45\textwidth]{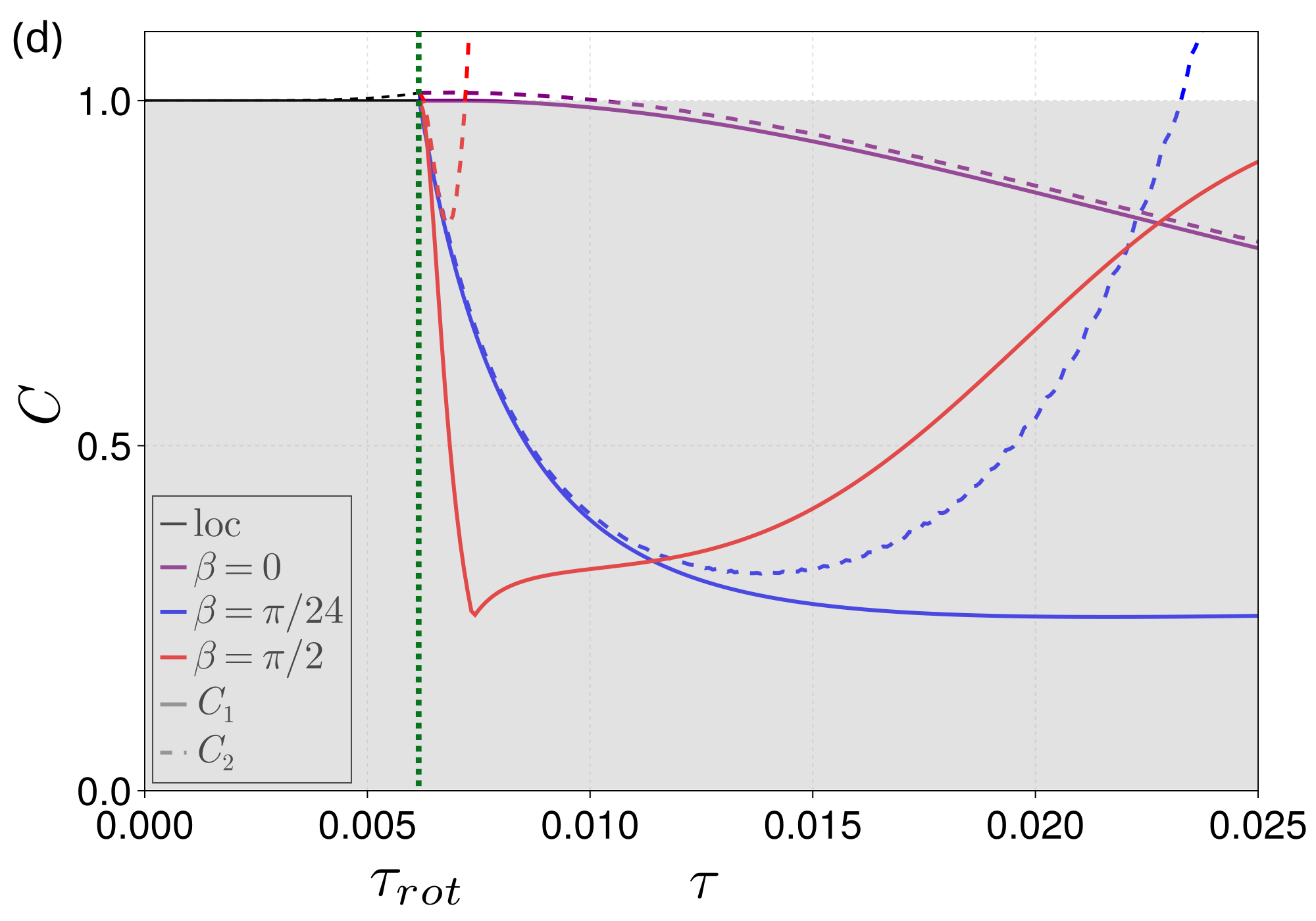}
\caption{{\bf Extension to a network of 3 and 4 nodes of the GID experiment.} (a) Evolution of the local (solid) and collective (dashed) squeezing, for $M=3$, $N= 1000$, following the protocol described in Fig ~\ref{Fig2}. Different colors correspond to different orientations of the local spin squeezing. Insets show the local squeezing angle for the different colors, and the geometric configuration for the $M=3$ network. (b) Evolution of the entanglement criteria ${\cal C}_{1}$ (solid) and ${\cal C}_{2}$ (dashed) following the same protocol. (c-d) Same figures as (a-b), but with $M=4$ BECs.}
\label{FigG}
\end{figure*}

In this section, we show and discuss the results for the GID experiment for $M=3$ and $4$, following the same protocol as shown in Fig.~3.
We show in Fig. ~\ref{FigG} (a) (resp. (c)) the evolution of the local and collective spin squeezing for various rotation angles $\beta$ for $M=3$ (resp. $M=4$). As in the $M=2$ case, the local squeezing parameter increases sharply after $\tau_{\rm rot}$ for $\beta = \pi/2$, while it remains almost constant for $\beta = 0$. On the other hand, for the collective squeezing, we see that it decreases slightly for $\beta = 0$ (this effect is stronger when we increase $M$), and we reach an important minimum (deeper the larger $M$) for $\beta = \pi/24$. We also observe a small decrease in the collective squeezing for $\beta = \pi/2$ before a sharp increase (which is due to the depolarization of the local spins).
All these observations correspond to what we see in Fig.  ~\ref{FigG} (b) (resp. (d)) for the evolution of the entanglement criteria ${\cal C}_{1}$ and ${\cal C}_{2}$ for $M=3$ (resp. $M=4$).

Again, the dynamics is much faster for $\beta = \pi/2$, and in that case criterion ${\cal C}_{2}$ (red dashed line) can only detect bipartite entanglement at short times after the beginning of the interaction. The dynamics is the slowest for $\beta = 0$ (solid and dashed purple lines), but we can detect bipartite entanglement provided that we wait for a long enough time (increasing $M$ still leads to a speedup of the dynamics). Finally, it is possible to clearly detect entanglement with criterion ${\cal C}_{2}$ for $\beta = \pi/24$, as we have a deep and broad minimum. It is worth noticing that the minimum gets deeper but thinner as we increase $M$, while it only gets deeper when using the stronger criterion ${\cal C}_{1}$.

\begin{thebibliography}{100}
\bibitem{note2}
In the case of $M=3$ and $4$, due to the anisotropy of dipolar interaction, it is important to have 3 subsystems forming an equilateral triangle in the $xy$ plane, while the fourth subsystem does not form a perfect tetrahedron, but is slightly closer to the other nodes in order to compensate for the $1-3 \cos^{2} \Theta$ dependence of the dipolar interaction ($\Theta$ being here the angle between the $z$ axis and the line drawn by the two nodes).
\bibitem{diffeo}
That is, no boundary-preserving diffeomorphism can map between configurations of different energies without altering the overall physical state -- including the gravitational field itself.
\end{thebibliography}
\end{document}